\documentclass{SCIS2026}
\makeatletter
\newcommand{\authorbreak}{%
	\g@addto@macro\@author{\\}%
	\g@addto@macro\@authortemp{\\}%
}

\begin{document}
	%%%%%%%%%%%%%%%%%%%%%%%%%%%%%%%%%%%%%%%%%%%%%%%%%%%%%%%
	%%% Authors do not modify the information below
	%%%
	\ArticleType{Review}
	%\SpecialTopic{}
	\Year{2025}
	\Month{January}
	\Vol{68}
	\No{1}
	\DOI{}
	\ArtNo{}
	\ReceiveDate{}
	\ReviseDate{}
	\AcceptDate{}
	\OnlineDate{}
	\AuthorMark{}
	\AuthorCitation{}
	%%%%%%%%%%%%%%%%%%%%%%%%%%%%%%%%%%%%%%%%%%%%%%%%%%%%%%%
	
	\title{Media Meets Communication in 6G: Fundamentals, Key Technologies, and Applications}%{XXXX}
	
	\author{Bingyan Xie$^1$, Longyu Zhou$^2$, Zihan Chen$^2$, Shunpu Tang$^3$, Mingyang Shi$^1$, Yu Tian$^1$, Guo Lu$^1$,\\ Yongpeng Wu$^1$, Tianhao Liang$^4$, Tony Q.S. Quek$^2$, Guangtao Zhai$^1$, Wenjun Zhang$^{1*}$}{}
	
	\address[1]{School of Information Science and Electronic Engineering, Shanghai Jiao Tong University, Shanghai, China\\}
	\address[2]{ISTD Pillar, Singapore University of Technology and Design, 8 Somapah Rd, Singapore\\}
	\address[3]{College of Information Science and Electronic Engineering, Zhejiang University, Hangzhou, China\\}	
	\address[4]{School of Information Science and Technology, Harbin Institute of Technology (Shenzhen), Shenzhen 518055, China}
	
	%%% Abstract. 
	\abstract{The rapid advancement of sixth-generation (6G) networks is accelerating the convergence of media intelligence and communication intelligence, driving media communication beyond conventional bit-level delivery toward intelligent, semantic-aware, and generative paradigms. Emerging media services require not only high data rates and low latency, but also semantic awareness, perceptual quality assurance, adaptive resource orchestration, trustworthy content processing, and personalized media generation. Meanwhile, media technologies are evolving from handcrafted signal processing and conventional coding toward artificial intelligence (AI)-driven representation learning, content understanding, and generative reconstruction. Motivated by these trends, this paper presents a systematic survey of media communication technologies for 6G vision communication by revisiting the evolution of communication and media technologies and clarifying the intrinsic relationship between media content processing and wireless transmission. We introduce a unified framework consisting of four key dimensions: AI-driven media technologies, media-aware wireless transmission, large model-enabled media communication, and intelligent network infrastructures. Specifically, AI-driven media technologies encompass media coding, content understanding, quality assessment, security and compliance detection, and AIGC-enabled media generation, while media-aware wireless transmission is examined from three complementary perspectives: semantic joint source-channel optimization, which jointly encodes task-relevant semantic information; source-aware transmission optimization, which leverages media characteristics for channel adaptation, prediction, and compensation; and channel-aware source optimization, which adapts media coding and reconstruction based on real-time channel conditions. Furthermore, large models enable multi-modal, token-based, and generative media communication by transforming dense media signals into compact representations and generative conditions. Intelligent network infrastructures, such as intelligent broadcasting networks, AI-RAN, edge caching, network slicing, and end-edge-cloud collaboration, further provide system-level support for scalable media service orchestration. Finally, we discuss key research challenges and future directions, including unified semantic representation and transmission protocols, collaborative media generation across distributed systems, large model-driven adaptive services, security and privacy protection, and standardization with experimental validation. Through this survey, we aim to provide a comprehensive understanding of how media processing, wireless transmission, generative intelligence, and network infrastructures can be jointly designed. These insights can support efficient, personalized, and trustworthy media services in future 6G networks.	
	}
	
	%%% Keywords.
	
	\maketitle
	\section{Introduction}
	\subsection{Research Background}
	\subsubsection{Evolution from 5G to 6G Communication Technologies}
	
	The evolution from 5G to 6G mobile communication is not simply a matter of achieving higher data rates, lower latency, or denser connections. More importantly, it reflects a change in how wireless systems are designed and what they are expected to support. In the 5G era, most technical advances were driven by stronger wireless air-interface technologies and the use of wider spectrum resources. For example, massive multiple-input multiple-output (MIMO) \cite{larsson2014massive}, millimeter-wave (mmWave) communication \cite{rappaport2013millimeter}, and flexible radio access were introduced to improve spectral efficiency, increase network capacity, and enhance the user-experienced data rate. Based on this technical foundation, the International Telecommunication Union (ITU) defined three representative 5G usage scenarios: enhanced mobile broadband (eMBB), ultra-reliable and low-latency communications (URLLC), and massive machine-type communications (mMTC). These scenarios mainly target high-rate broadband access, low-latency industrial control, and large-scale Internet-of-Things (IoT) connectivity, respectively \cite{itur2015imt2020}.
	
	Although 5G has greatly improved wireless transmission capability, its design philosophy is still largely centered on connectivity and bit-level data delivery. In other words, 5G has made it possible to transmit data faster, more reliably, and to more devices, but the network remains largely unaware of the content semantics and service intent of transmitted data. This limitation becomes more evident in emerging applications such as immersive media, digital twins, autonomous systems, and distributed artificial intelligence (AI). In these applications, communication is no longer an isolated process of delivering bitstreams. Instead, it is closely related to computing, sensing, learning, and media processing. Therefore, future networks need to support not only higher bandwidth and lower latency, but also a deeper understanding of content, tasks, user experience, and service requirements.
	
	Compared with 5G, 6G is expected to evolve from a connectivity-centric network into an intelligence-native and service-oriented infrastructure. According to the ITU-R IMT-2030 framework, IMT-2030 is envisioned to include six usage scenarios: immersive communication, hyper-reliable and low-latency communication, massive communication, ubiquitous connectivity, AI and communication, and integrated sensing and communication \cite{itur2023imt2030}. This evolution shows that 6G will no longer be limited to connecting devices and delivering information between endpoints. Instead, it will combine communication with computing, sensing, intelligence, and security in a unified system. In this system, AI is not only used as an external tool to optimize network operation, but also becomes a native capability embedded in the network architecture \cite{letaief2019roadmap, cui2025overview}.
	
	This transition is particularly important for media-related services. Future 6G networks are expected to support various media-centric services, e.g. extended reality (XR), digital twin synchronization, and AI-generated media services. These services are very different from conventional video streaming. They are more interactive, more sensitive to delay, and more dependent on real-time computing and intelligent content processing. For example, an XR user may need the network to deliver visual content with low latency, adapt the media quality to the wireless channel, and support real-time rendering at the edge. Similarly, a digital twin system may require continuous synchronization between physical scenes and virtual models. Therefore, the communication system is gradually changing from a passive bit-pipe into an active media-service platform. It needs to coordinate compression, transmission, and generation across end-edge-cloud systems.
	
	In this context, the objective of communication is shifting from reliable bit delivery toward reliable content delivery. Although traditional indicators, such as throughput, latency, and block error rate, remain essential, they are insufficient to characterize the quality of future media-oriented communication services. For visual and immersive services, the received content should not only be decoded correctly at the bit level, but also preserve important semantics, perceptual quality, and task utility for personalized user experiences. This motivates cooperative design between media representation and wireless transmission. Such a trend also provides the foundation for semantic communication and generative media communication, where the network transmits task-relevant semantics, generative prompts, and compact tokens for efficient and adaptive media delivery \cite{strinati2021beyond,zhang2026generative}.
	
	\subsubsection{Evolution from 5G to 6G Media Technologies}
	
	In parallel with the evolution of wireless networks, media technologies are also moving toward a new stage. In the 4G and 5G eras, mobile media services were mainly driven by the rapid growth of video traffic, including high-definition (HD) streaming, short videos, live broadcasting, mobile gaming, and social media applications. According to Cisco’s VNI forecast for 2017–2022, video traffic was projected to account for over 80$\%$ of global Internet traffic by 2022, indicating that video has become one of the dominant services in modern networks \cite{barnett2018cisco}. To support this demand, media technologies primarily focused on improving compression efficiency and visual quality under limited bandwidth.
	
	Early media processing systems were mainly built upon manually designed modules. In video coding, standards such as H.264 \cite{wiegand2003overview}, high efficiency video coding (HEVC) \cite{sullivan2012overview}, and versatile video coding (VVC) \cite{bross2021overview} have achieved continuous compression gains through carefully designed modules, including prediction, transform, quantization, filtering, and entropy coding. Similar ideas were also widely adopted in low-level vision tasks, such as denoising, super-resolution, frame interpolation, and image enhancement. These methods are reliable and interpretable, but they depend heavily on handcrafted priors. As visual content becomes more diverse and complex, it becomes increasingly difficult to describe its spatial, temporal, and perceptual characteristics using manually designed models alone.
	
	The development of deep learning has changed the way media signals are represented and processed. Instead of relying only on explicit rules, neural networks can learn visual representations directly from data. This has promoted rapid progress in learned image compression \cite{balle2017end}, learned video compression \cite{lu2019dvc}, image restoration, frame interpolation, and perceptual quality assessment. Compared with traditional methods, deep models are better at capturing complex textures, motion patterns, and spatio-temporal dependencies. Therefore, media processing is gradually evolving from handcrafted pipelines to data-driven and content-adaptive systems.
	
	More recently, generative models and multi-modal large language models (MLLMs) have further shifted media technologies from low-level signal processing toward semantic understanding and content generation. Generative adversarial networks (GANs) \cite{goodfellow2014generative} and diffusion models \cite{ho2020denoising} can synthesize realistic visual content, recover missing details, and generate high-quality images or videos from compact conditions. Meanwhile, vision-language models (VLMs), such as CLIP \cite{radford2021learning}, BLIP \cite{li2022blip}, Flamingo \cite{alayrac2022flamingo}, and LLaVA \cite{liu2024visual}, enable media systems to connect visual signals with semantic concepts and natural language instructions. With these models, visual content can be described, searched, edited, and generated at the semantic level.
	
	This evolution gives rise to content-aware media technologies. Media is no longer regarded only as pixels to be compressed and reconstructed. Instead, the system can understand the content, identify the information that matters most, and adapt its representation according to the task, user, and network condition. For example, under limited bandwidth, it may be more efficient to transmit a low-quality reference, semantic tokens, textual descriptions, or generative prompts, and then use a generative model at the receiver to reconstruct the desired content. This idea is consistent with generative video communication, where traditional video transmission provides reliability and controllability, while generative models improve expressiveness and perceptual quality \cite{zhang2026generative}.
	
	Therefore, the evolution from 5G to 6G media technologies can be viewed as a transition from handcrafted low-level processing to deep-learning-based media intelligence, and further to semantic-aware and generative media representation. This transition naturally motivates a closer integration between media processing and wireless communication. Future media services will depend not only on how many bits are transmitted, but also on what content is transmitted, how its semantics are preserved, and how it is reconstructed or generated at the receiver.

	\begin{figure*}[htbp]
	\centering
	\includegraphics[width=6.7in]{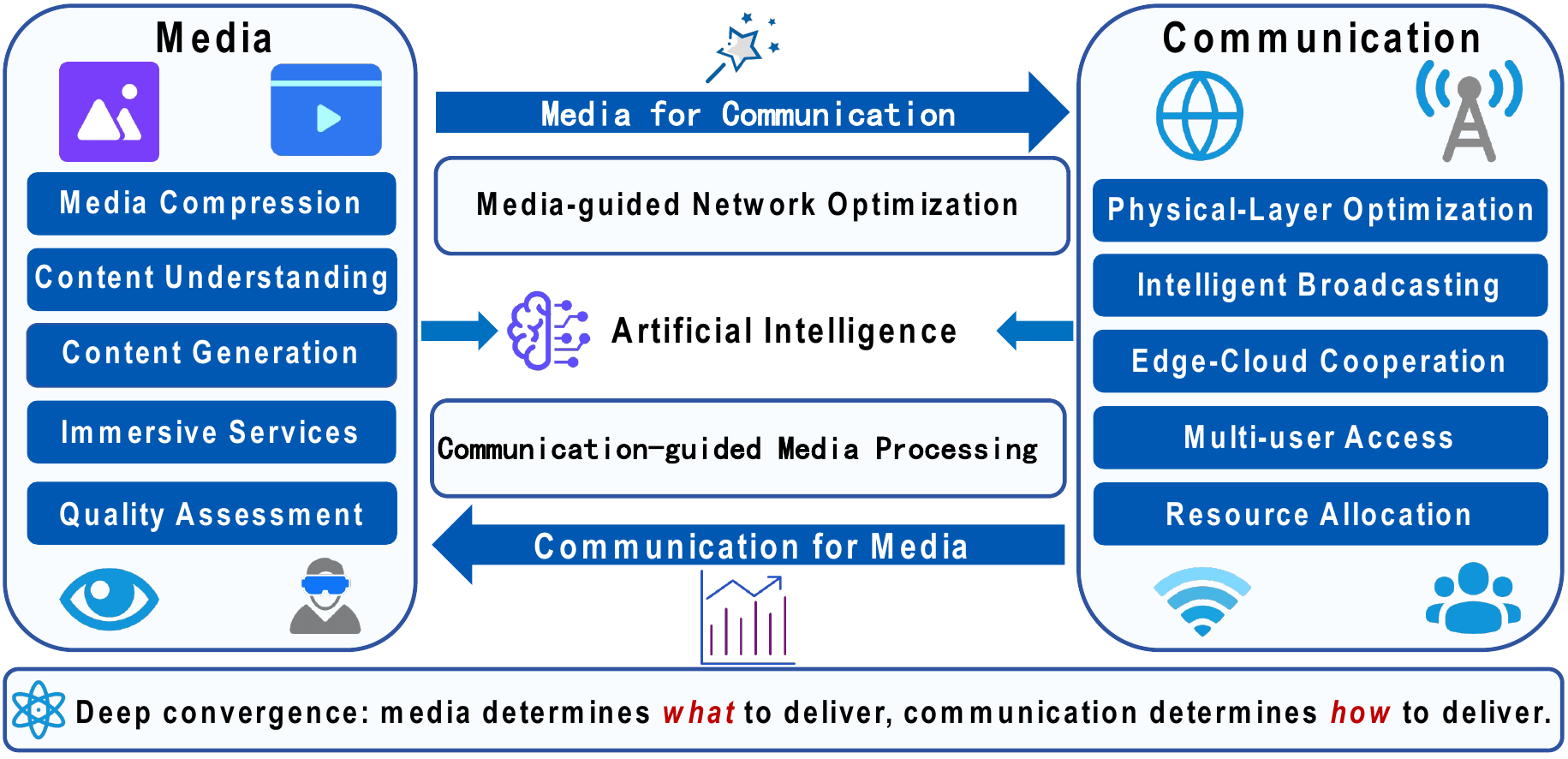}
	\caption{Relationship between media and communication.}
	\label{fig_1}
	\end{figure*}

	\subsubsection{Intrinsic Relationship between Media and Communication}
	
	Media and communication are intrinsically coupled. As illustrated in Fig. \ref{fig_1}, media determines what information should be delivered, while communication determines how such information can be delivered under wireless and computational constraints. In conventional systems, these two parts are usually designed separately. A source codec first compresses visual signals into bitstreams, and the communication system then transmits these bitstreams as reliably as possible. This modular design has supported the success of modern video services for decades. However, it becomes less sufficient for 6G media applications, where real-time interaction, semantic understanding, and personalized content generation must be supported in highly dynamic wireless environments.
	
	This limitation can also be understood from a broader communication perspective. Classical information theory mainly addresses the technical problem of reproducing transmitted symbols accurately. However, Shannon and Weaver pointed out that communication also involves semantic and effectiveness levels, namely whether the transmitted message preserves its intended meaning and achieves its desired effect \cite{shannon1949mathematical}. This view is particularly relevant to vision communication, which is not only related to bits but also visual content. For users, the quality of a received video is not determined only by bit-level correctness. It also depends on whether the reconstructed content conveys the intended scene, supports the target task, and provides a satisfactory perceptual experience.
	
	This observation has motivated increasing interest in semantic and goal-oriented communication \cite{strinati2021beyond, gunduz2023beyond}. Unlike conventional communication, which treats all bits or symbols uniformly for reliable delivery, semantic communication focuses on the meaning and task-relevant information carried by the source. This is especially suitable for visual media. In an image or video, different content elements do not contribute equally to human perception or machine analysis. For example, a moving object, a human face, or an interaction region may be much more important than a smooth background. Therefore, media communication should not only focus on how to transmit fewer bits, but also which information is worth protecting and how it contributes to the final service quality.
	
	The coupling between media and communication becomes even stronger in wireless environments. On one hand, media content can guide network operation. Motion intensity, regions of interest, and semantic priority can help the network jointly allocate communication, computing, and caching resources more effectively. On the other hand, wireless conditions can guide media processing. When the channel is stable and bandwidth is sufficient, the system may transmit high-fidelity video streams. When the channel becomes poor, it may reduce the resolution, protect key semantic regions, transmit compact tokens, or rely on generative models to recover missing details at the receiver. In this way, media representation and wireless transmission form a closed loop rather than two isolated stages.
	
	In 6G, this relationship is expected to evolve from simple adaptation to deep convergence. Future media communication systems will jointly consider the complete media transmission system with source coding, channel coding, transmission scheduling, edge computing, generative reconstruction, and quality evaluation. The role of the receiver will extend beyond bitstream decoding, enabling semantic understanding, content enhancement, and media generation based on the received cues and local context. This vision is consistent with the overall structure of our survey, where AI-driven media processing, media-aware wireless transmission, large model-enabled generative communication, and media-oriented intelligent networks form the key pillars of 6G vision communication. Therefore, the central question is no longer simply how to transmit media over communication networks, but how to build an integrated vision communication system in which media intelligence and network intelligence work together.
	
	\subsection{Research Outline}
	
	\subsubsection{Vision of Convergence}
	
	Our vision of media-communication convergence is to build a visual communication paradigm in which media processing and wireless transmission are no longer treated as independent procedures. Instead, visual content, user demand, wireless channel state, edge computing capability, and generative models are jointly considered. Under this vision, the network does not merely deliver compressed bitstreams; it actively supports robust transmission, on-demand content delivery, personalized generation, and media-oriented network services.
	
	\paragraph{Robust Vision Transmission.}
	Future visual communication should maintain stable perceptual and semantic quality under dynamic wireless channels. Important visual information, such as key frames, salient regions, and task-relevant semantics, can be prioritized, protected, or reconstructed to avoid severe quality degradation.
	
	\paragraph{On-demand Content Transmission.}
	Visual content should be transmitted according to user needs, device capabilities, and service requirements. Instead of delivering all pixels uniformly, the system may transmit full video, regions of interest, semantic features, visual tokens, or compact descriptions depending on the application.
	
	\paragraph{Personalized Generative Transmission.}
	Generative models enable media delivery from compact cues, such as prompts, low-resolution references, motion hints, or semantic tokens. The receiver can reconstruct or generate content adapted to user preference, terminal type, and network condition.
	
	\paragraph{Network Infrastructures for Vision Communication.}
	6G networks should provide native support for visual services through edge computing, intelligent caching, network slicing, artificial intelligence radio access network (AI-RAN), and end-edge-cloud collaboration. The network will evolve from a bit-delivery pipe into an intelligent platform for media processing, transmission, and generation.
	
	\begin{table*}[htbp]
		\centering
		\caption{Comparison of recent surveys related to 6G+AI, semantic communication, and generative AI for communication.}
		\label{tab1}
		\renewcommand{\arraystretch}{1.16}
		\setlength{\tabcolsep}{2.6pt}
		\scriptsize
		\begin{tabular}{
				>{\raggedright\arraybackslash}p{2.2cm}
				>{\raggedright\arraybackslash}p{2.8cm}
				>{\raggedright\arraybackslash}p{5.4cm}
				>{\raggedright\arraybackslash}p{5.6cm}
			}
			\toprule
			\textbf{Survey} 
			& \textbf{Main Focus} 
			& \textbf{Main Content} 
			& \textbf{Contributions} \\
			\midrule
			
			Cui \textit{et al.} \cite{cui2025overview}
			& AI and communication for 6G
			& Reviews the integration of AI and 6G communications from three perspectives: AI for network, network for AI, and AI as a service. It further discusses enabling technologies, representative scenarios, standardization progress, and future research opportunities.
			& Provides a comprehensive framework for understanding how AI enhances wireless networks and how 6G networks support distributed AI services, offering a broad reference for AI-native 6G system design. \\
			
			\midrule
			
			Das \textit{et al.} \cite{das2025comprehensive}
			& Emerging AI technologies for 6G
			& Surveys representative AI techniques and their applications in 6G networks, covering intelligent resource allocation, security, reliability, optimization, and network automation.
			& Summarizes recent AI-driven research trends for 6G communications and identifies major challenges and opportunities for deploying AI in future wireless systems. \\
			
			\midrule
			
			Khan and Schmid \cite{khan2024airan}
			& AI-RAN in 6G
			& Reviews AI-RAN architectures, use cases, and open challenges in 6G, including radio-frequency (RF) planning, capacity and coverage optimization, mobility management, handover, and radio resource control.
			& Provides a dedicated taxonomy for AI-RAN and clarifies how AI/machine learning (ML), edge computing, and radio access networks can be jointly designed to improve automation, adaptability, and network intelligence. \\
			
			\midrule
			
			Lu \textit{et al.} \cite{lu2024semantics}
			& Semantics-empowered communication
			& Provides a tutorial-style survey of semantic communication, including theoretical foundations, performance metrics, datasets, toolkits, enabling techniques, and practical implementations.
			& Establishes a systematic taxonomy for semantic communication and clarifies the transition from conventional bit-level transmission to meaning-oriented communication. \\
			
			\midrule
			
			Guo \textit{et al.} \cite{guo2025semantic}
			& Semantic communication networks
			& Reviews semantic communication networks from the perspectives of system architecture, semantic interaction, security, privacy, and multi-agent intelligence.
			& Extends semantic communication from point-to-point links to networked semantic systems, highlighting key issues in semantic networking, trustworthiness, and privacy protection. \\
			
			\midrule
			
			Shen \textit{et al.} \cite{shen2023toward}
			& Immersive communication in 6G
			& Surveys immersive 6G services, including XR, holographic communication, haptic interaction, and remote multi-sensory applications, with emphasis on service requirements and network challenges.
			& Identifies key communication requirements for immersive services, such as high data rate, low latency, high reliability, synchronization, and quality-of-experience assurance. \\
			
			\midrule
			
			Xu \textit{et al.} \cite{xu2024unleashing}
			& Edge-cloud generative AI in mobile networks
			& Reviews AI-generated content (AIGC) services in mobile networks, covering generative model lifecycle management, cloud-edge-mobile collaboration, service deployment, privacy, and security.
			& Provides a systematic view of how mobile networks can support real-time, personalized, and scalable AIGC services through collaborative computing and communication. \\
			
			\midrule
			
			Van Huynh \textit{et al.} \cite{vanhuynh2024generative}
			& Generative AI for wireless communications
			& Surveys the applications of generative AI in wireless physical-layer communications, including signal classification, channel estimation, equalization, RIS-aided transmission, and joint source-channel coding.
			& Demonstrates the potential of generative AI in modeling complex wireless signal distributions and improving physical-layer design, optimization, and robustness. \\
			
			\midrule
			
			Oukebdane and Shah \cite{oukebdane2025computer}
			& Computer vision-powered 6G networks
			& Reviews the role of computer vision in 6G wireless networks, including visual perception, context awareness, intelligent monitoring, environment understanding, and related applications.
			& Highlights how visual perception can support intelligent and context-aware 6G networks, providing insights into vision-assisted network management and optimization. \\
			
			\midrule
			
			\textbf{Ours}
			& \textbf{Vision communication in 6G}
			& \textbf{Systematically reviews 6G vision communication from AI-driven media coding, understanding, and generation to media-aware wireless transmission, large model-enabled token/generative communication, and intelligent networks for media applications.}
			& \textbf{Provides a comprehensive synthesis of 6G vision communication, clarifying its technical evolution, core research themes, representative applications, and future challenges toward media communication convergence.} \\
			
			\bottomrule
		\end{tabular}
	\end{table*}
	
	However, as summarized in Tab. \ref{tab1}, existing 6G-related surveys have not yet provided a dedicated and systematic review of the fundamentals, key technologies, and applications of vision communication. The main contributions of this paper are summarized as follows:
	
	\begin{itemize}
		\item \textbf{Detailed overview of AI-driven media technologies:} 
		We present an up-to-date overview of AI-driven media coding, understanding, and generation, which provides the technical foundation for media-communication convergence in 6G vision communication.
		
		\item \textbf{Comprehensive taxonomy of media-communication interaction:} 
		We develop a comprehensive taxonomy to characterize the interaction between media and communication, highlighting how source characteristics and channel conditions can be jointly exploited for efficient, robust, and task-oriented transmission.
		
		\item \textbf{Cross-layer architecture for media services:} 
		We analyze cross-layer network architectures that support emerging media services, including intelligent broadcasting networks, edge-assisted media processing, and media-centric AI-RAN, with emphasis on coordinated communication, computing, caching, and intelligence.
		
		\item \textbf{Review of emerging paradigms:} 
		We review new paradigms for vision communication, including generative media communication, multi-modal media transmission, and large model-enabled token-based communication. These paradigms open new opportunities for semantic-aware, personalized, and low-bitrate visual services.
		
		\item \textbf{Challenges and future outlook:} 
		We identify key open challenges, such as low-latency delivery, compatibility with existing communication systems, personal data security, controllable generation, and content-oriented quality assessment. We further outline promising research directions toward practical 6G vision communication.
	\end{itemize}
	
	\begin{figure*}[htbp]
		\centering
		\includegraphics[width=6.7in]{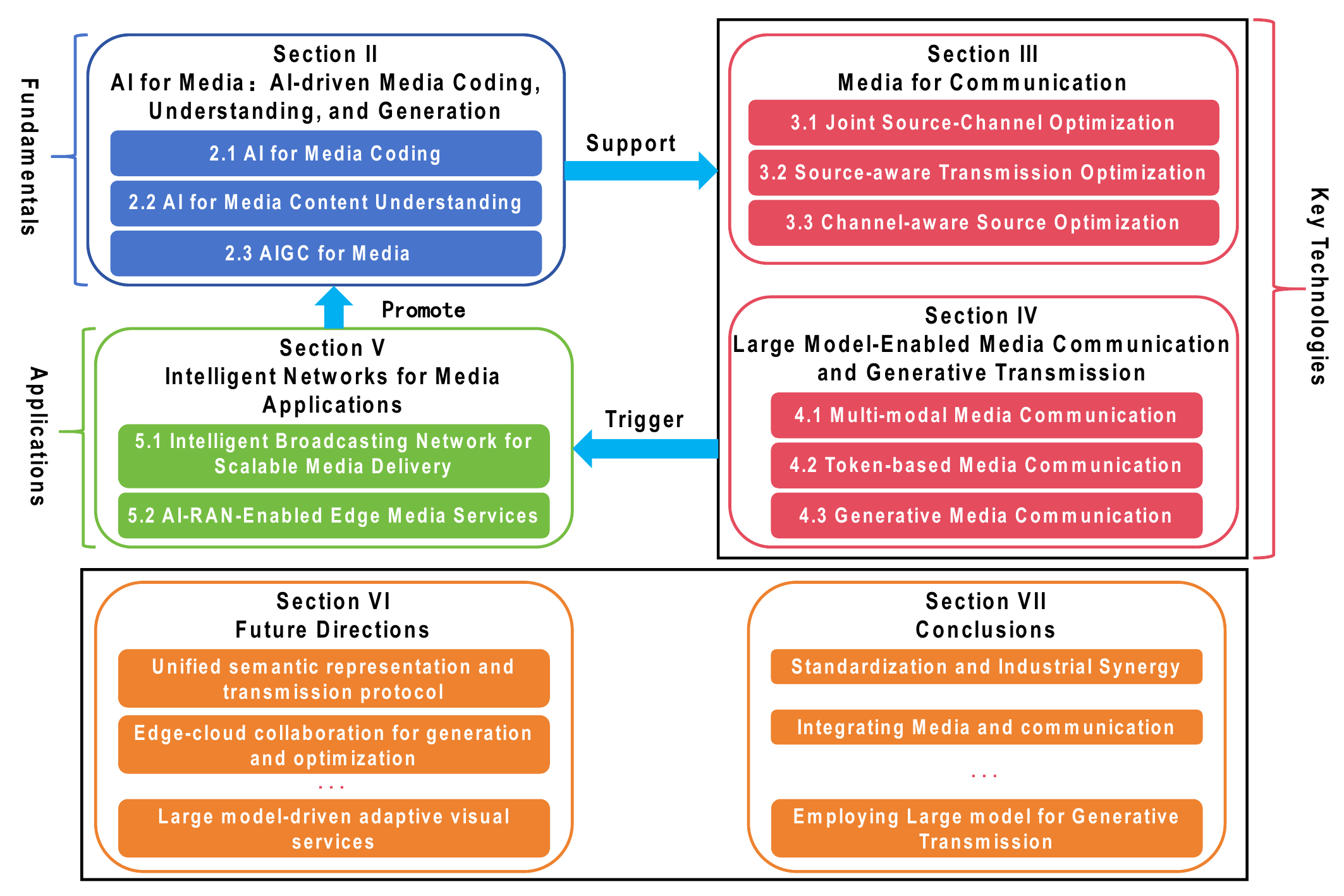}
		\caption{Paper organizations and relationships among sections.}
		\label{fig_2}
	\end{figure*}
	
	\subsubsection{Paper Organization}
	
	The remainder of this paper is organized as follows. Section 2 reviews AI-driven media technologies, including media coding, media content understanding, and AIGC for media. Section 3 discusses media-communication interaction from the perspectives of joint source-channel optimization, source-aware transmission optimization, and channel-aware source optimization. Section 4 presents emerging paradigms enabled by large models, including multi-modal media communication, token-based media communication, and generative media communication. Section 5 investigates intelligent network architectures for media applications, covering intelligent broadcasting networks and AI-RAN-enabled edge media services. Section 6 outlines future research directions for 6G vision communication. Finally, Section 7 concludes this paper. The structure of this paper is shown in Fig. \ref{fig_2}.
	
	\section{AI-driven Media Coding, Understanding, and Generation}

\begin{figure*}[htbp]
	\centering
	\includegraphics[width=6.7in]{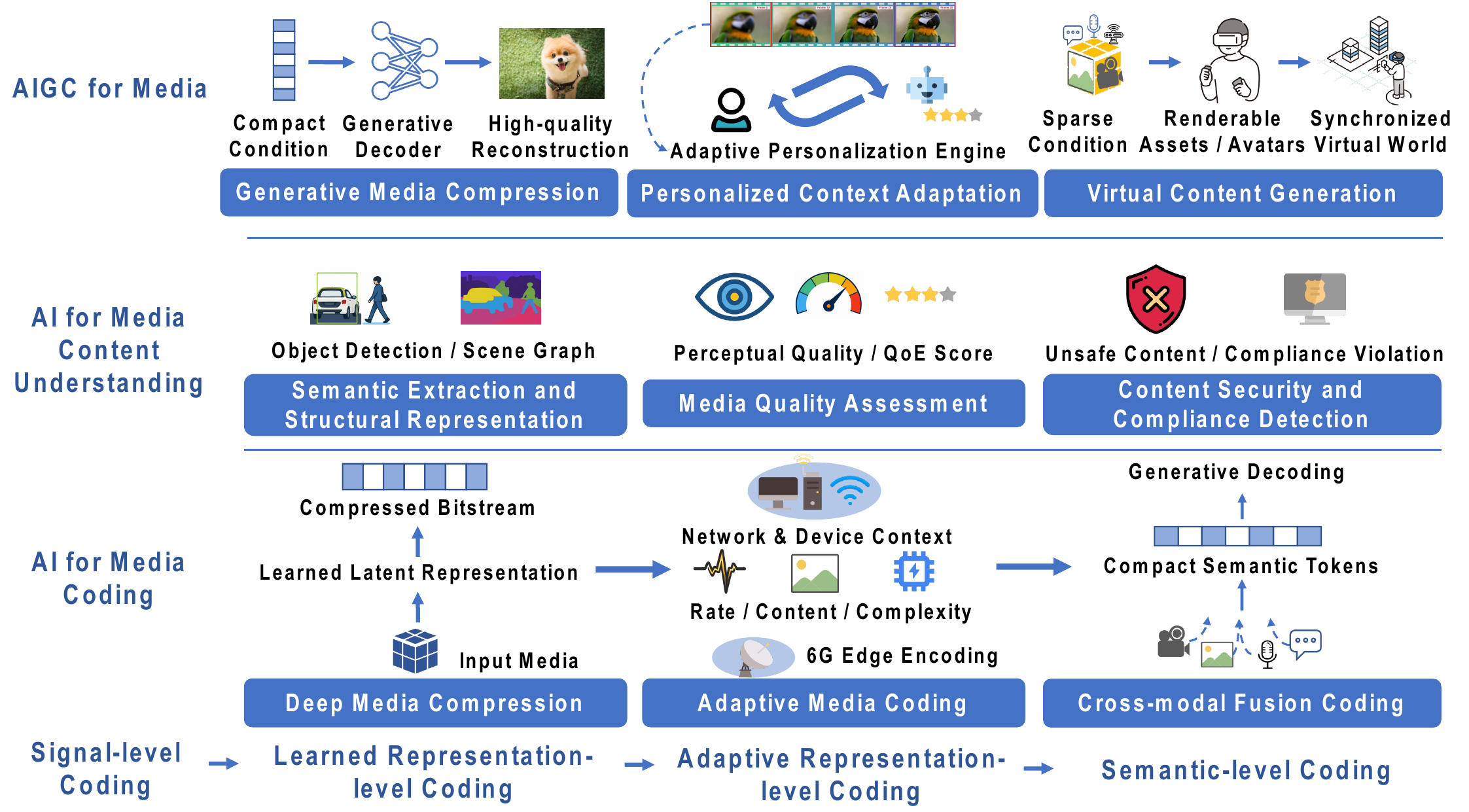}
	\caption{Overview of AI-driven media coding, understanding, and generation.}
	\label{fig_3}
	\end{figure*}

	This section focuses on AI-driven media technologies that form the source-side foundation of 6G vision communication. Deep learning, foundation models, and generative AI are reshaping media coding, content understanding, quality assessment, security detection, and generation-guided reconstruction. These capabilities produce compact representations, structured semantics, perceptual feedback, and generative priors that can later be exploited by wireless transmission and network orchestration. The structure of Section 2 is shown in Fig. \ref{fig_3}.

	\subsection{AI for Media Coding}
	
	Media compression has long been the foundation of modern visual communication systems. Conventional standards, such as H.264/AVC, HEVC, and H.266/VVC, have achieved high coding efficiency through optimized block partitioning, motion compensation, transform coding, quantization, and entropy coding. However, these hybrid architectures rely on manually designed signal-processing modules and mainly target pixel-level redundancy reduction. Under low-bitrate, real-time, or edge-constrained operation, conventional codecs often suffer from texture degradation, structural distortion, and perceptual quality loss, while their computational complexity challenges real-time and edge-oriented deployment. These limitations motivate the transition from handcrafted signal representations toward learned media representations.
	
	\begin{figure*}[htbp]
	\centering
	\includegraphics[width=6.3in]{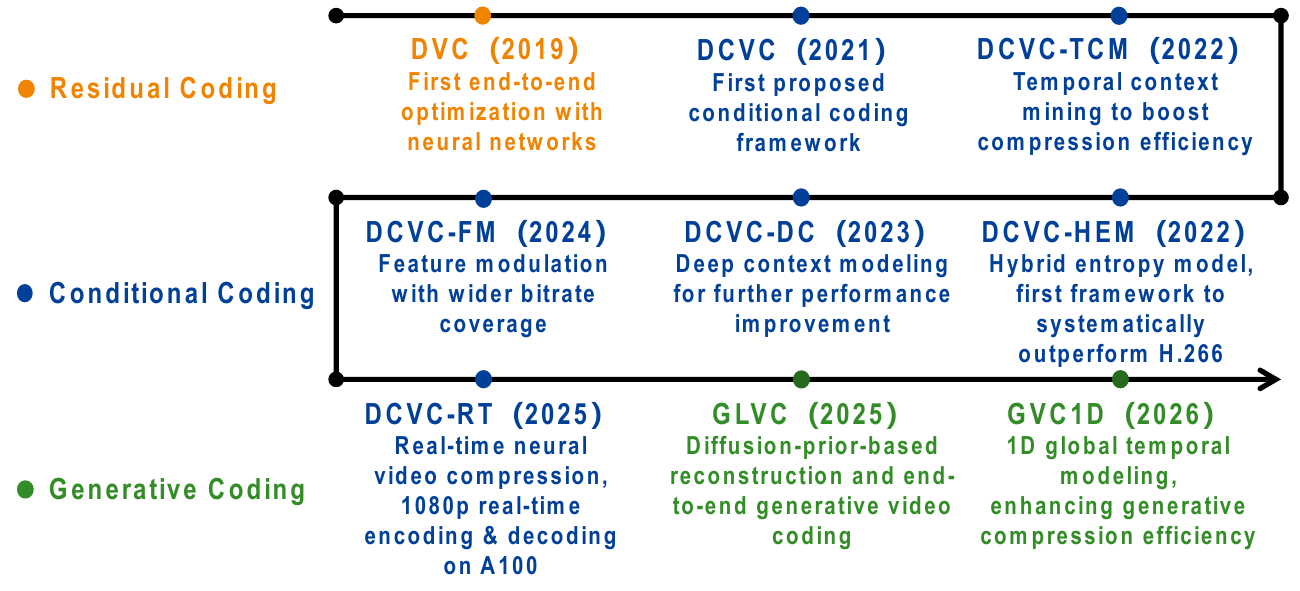}
	\caption{Evolutions of deep video coding.}
	\label{fig_4}
	\end{figure*}
	
	\subsubsection{Deep Media Compression}

	Deep media compression advances this transition by learning compact latent representations directly from data. Instead of relying on handcrafted prediction, transform, quantization, and entropy-coding modules, neural codecs jointly optimize representation learning, entropy modeling, and reconstruction. This changes compression from efficient pixel coding to learned representation for communication.

	From the perspective of representation learning, deep video compression has evolved through three main stages, as illustrated in Fig.~\ref{fig_4}: residual representation learning, contextual representation learning, and unified semantic representation learning. The first stage follows the \textbf{conventional predictive coding paradigm}, where early neural codecs mainly replace individual coding modules with learned networks while still transmitting prediction residuals. A representative example is Deep Video Compression (DVC)~\cite{lu2019dvc}, which jointly optimizes motion estimation, motion compensation, residual coding, and entropy modeling in an end-to-end framework. The second stage moves toward \textbf{contextual representation learning}, where spatial context, temporal dependencies, and latent priors are jointly exploited to improve compression efficiency. Deep Contextual Video Compression (DCVC)~\cite{li2021dcvc} introduces context-aware prediction and conditional entropy modeling, while lightweight extensions such as DCVC-RT~\cite{jia2025dcvcrt} further reduce computational complexity for real-time deployment. The third stage shifts toward \textbf{unified semantic representation learning}, where multiple coding tasks are supported within a shared latent space. Unified Image and Intra-frame Compression (UI2C)~\cite{xiang2025ui2c} jointly optimizes image compression and intra-frame video coding, while Unified Learned Video Compression (UniLVC)~\cite{zhang2026unilvc} integrates motion modeling, context propagation, and rate-distortion optimization within a Transformer-based architecture.

	Deep media compression is therefore moving from residual-based coding to contextual modeling and unified representation learning. As neural networks become the core mechanism for media representation, Transformer architectures, multi-modal foundation models, diffusion priors, and generative models can further connect compression, understanding, and reconstruction for generation-aware media transmission.
	
	\subsubsection{Adaptive Media Coding}

	While deep media compression improves learned media representations, future 6G media services require codecs to operate under rapidly changing bandwidth, content complexity, device capability, and computational budgets. A fixed coding strategy is difficult to generalize across such conditions. Consequently, media coding is evolving from static representation learning toward adaptive representation learning, where coding decisions are dynamically optimized according to network conditions, media characteristics, and system constraints.

	Recent studies on adaptive media coding can be broadly categorized into three directions: rate-aware adaptation, content-aware adaptation, and complexity-aware adaptation.

	\begin{itemize}
		\item \textbf{Rate-aware adaptation:} 
		Rate-aware adaptation aims to support flexible bitrate requirements with a single or scalable neural coding model. Since neural compression involves nonlinear relationships among latent representations, bitrate, and distortion, traditional rate-control mechanisms cannot be directly applied. Variable-rate neural codecs address this issue by using conditional autoencoders or gain-unit mechanisms to adjust latent representations according to target bitrates~\cite{choi2019variable,liu2024rdc}. Further studies model rate-distortion behavior in latent spaces for explicit rate control~\cite{xu2026feedback}, while reinforcement learning formulates bitrate allocation as a sequential decision problem for cross-frame rate-distortion optimization~\cite{cong2026reinforced}. These works indicate that bitrate adaptation is shifting from handcrafted control rules toward learned rate-distortion optimization.
	
		\item \textbf{Content-aware adaptation:} 
		Content-aware adaptation adjusts coding strategies according to visual importance, motion dynamics, texture complexity, and semantic relevance. Since different regions and frames contribute unequally to perception and downstream tasks, uniform coding may lead to inefficient resource allocation. Content-weighted compression learns spatial importance maps to allocate more bits to perceptually important regions~\cite{li2023neural}, while saliency-aware and region-of-interest coding further exploit visual attention and semantic significance~\cite{yura2022roinvc}. Recent studies extend adaptation from bitrate allocation to representation learning, dynamically adjusting latent features according to scene complexity and spatio-temporal characteristics~\cite{chen2024sparsedense,tang2025canerv}. Task-oriented and semantic-aware coding further optimize representations for downstream tasks instead of pixel reconstruction alone~\cite{xiao2024task}.
	
		\item \textbf{Complexity-aware adaptation:} 
		Complexity-aware adaptation targets the deployment cost of neural codecs. Unlike conventional codecs with relatively stable computational complexity, neural codecs rely on deep architectures whose inference cost varies across devices and scenarios. Adaptive path-selection methods, such as AdaDPCC~\cite{zhang2025adadpcc}, dynamically adjust feature propagation paths according to scene complexity. Other studies explore complexity-scalable compression and dynamic inference, where only selected latent channels, network branches, or Transformer tokens are activated under limited computational budgets~\cite{wang2024complexity}. These approaches extend conventional rate-distortion optimization toward rate-distortion-complexity optimization, enabling neural codecs to better support heterogeneous 6G devices.
	\end{itemize}

	Adaptive media coding therefore spans network conditions, media content, and computational constraints. Modern adaptive codecs increasingly treat representation learning, transmission optimization, and resource scheduling as a unified problem, enabling more flexible media transmission for 6G vision communication.

	\subsubsection{Cross-modal Fusion Media Coding}

	Cross-modal fusion media coding aims to jointly represent and compress heterogeneous media sources, such as images, videos, audio, and text, within a unified representation space. Different from modality-specific coding schemes that independently encode and transmit each modality, cross-modal fusion coding exploits both intra-modal redundancy and inter-modal correlations. By modeling complementary information across modalities, it can reduce redundant transmission, improve coding efficiency, and preserve cross-modal consistency in reconstructed media content.

	Compared with conventional media coding, cross-modal fusion coding provides two main advantages. First, by mapping different modalities into a shared latent or embedding space, the system can exploit high-level correlations among visual, auditory, and textual signals. This is particularly useful for multi-modal services, where different media streams may describe the same scene, event, or user interaction from different perspectives. Second, joint representation enables the encoder to allocate coding resources according to modality importance, content complexity, and service requirements. Instead of uniformly preserving all pixel, waveform, or text details, the system can selectively encode the most informative components, thereby improving bitrate efficiency and robustness under bandwidth-constrained scenarios~\cite{xie2025msc}.

	Recent studies have extended cross-modal coding toward unified multi-modal representation and generative reconstruction. A unified coding framework can jointly represent images, text, audio, and video within a shared feature space, so that multi-modal information is no longer encoded as several isolated bitstreams but as compact representations that preserve cross-modal dependencies~\cite{mortaheb2025msc}. This paradigm is especially beneficial for low-bandwidth and resource-constrained environments, where transmitting all modality-specific details may be inefficient or unnecessary. In parallel, generative models further shift cross-modal coding from explicit signal reconstruction to model-assisted reconstruction~\cite{yin2026gvsc}. Instead of transmitting complete multi-modal signals, the encoder may send compact conditions, prompts, structural cues, or latent representations, while the receiver uses generative models to recover missing or highly compressed content.

	The rapid development of multi-modal foundation models further supports this trend. Recent models such as LanguageBind~\cite{zhu2024languagebind} and OneEncoder~\cite{faye2025oneencoder} show that images, videos, audio, and text can be aligned through compatible tokens or latent features within a unified embedding space. Such representations provide a foundation for future coding architectures in which compression, reconstruction, modality conversion, and content generation are jointly supported by shared multi-modal features rather than separate modality-specific codecs. Cross-modal fusion media coding is therefore moving from independent unimodal compression toward collaborative multi-modal representation and generative reconstruction, bridging efficient compression with intelligent media reconstruction in 6G media services.

	\subsection{AI for Media Content Understanding} 

	AI for media content understanding provides an important interface between visual sources and 6G vision communication systems. Different from conventional computer vision tasks that mainly focus on recognition accuracy, 6G-oriented media services require AI models to identify task-relevant content, evaluate media quality, and detect unsafe or non-compliant information before storage, transmission, or dissemination. This subsection reviews three key functions: semantic extraction and structural representation, media quality assessment, and content security and compliance detection. These functions transform raw visual streams into structured media information, thereby supporting content-aware coding, edge inference, secure distribution, and receiver-side reconstruction.

	\subsubsection{Semantic Extraction and Structural Representation}

	In 6G vision communication, semantic extraction provides a content interface between raw visual signals and media-aware transmission. The system must identify objects, regions, events, and reconstruction cues that are relevant to perception, interaction, or downstream analysis. Visual media should therefore be transformed into structured content representations that can guide coding, transmission protection, receiver-side enhancement, and generative reconstruction~\cite{cui2025overview,zhang2026generative}.

	This requirement is first reflected in object-level and region-level understanding. Detection and panoptic segmentation provide explicit spatial units for objects, regions, and scene categories~\cite{carion2020end}. Vision-language pre-training further aligns visual entities with natural-language concepts~\cite{radford2021learning}, while open-vocabulary grounding and promptable segmentation enable language-conditioned localization and interactive region parsing~\cite{liu2024grounding,kirillov2023segment,ravi2025sam}. Unified prompt-based models integrate detection, grounding, captioning, and segmentation within a common interface~\cite{xiao2024florence}. These methods make visual content explicitly selectable, allowing task-relevant objects, user-attended regions, redundant backgrounds, and privacy-sensitive areas to be distinguished before coding, transmission, or enhancement.

	Beyond object and region extraction, many 6G media services require structural and temporal representations. Scene graph methods describe visual content through object-attribute-relation structures~\cite{krishna2017visual,xu2017scene}, while panoptic and open-set scene graphs extend relation modeling to dense and open-world scenes~\cite{yang2023panoptic,zhou2024openpsg}. Dynamic scene graphs further represent spatio-temporal scene evolution~\cite{yang20234d}. In parallel, video representation learning and video foundation models provide compact temporal features for cross-frame modeling, video-text alignment, and predictive understanding~\cite{wang2024internvideo2,lin2024video,assran2025v}. Graph-based representations offer explicit scene structures, whereas video foundation models provide dense and generalizable temporal features; together, they support content-aware coding, edge inference, interactive control, and generative reconstruction.

	Although these methods differ in architecture, their common objective is to convert unstructured visual streams into compact, interpretable, and communication-friendly content representations. Such representations can support region-adaptive coding, semantic-priority transmission, privacy-preserving processing, and receiver-side reconstruction. However, representations produced by different models, such as semantic tokens, scene graphs, and vision-language embeddings, are not naturally aligned. Open-world models may also introduce semantic drift, domain instability, and privacy leakage. Therefore, semantic extraction should be designed not only for recognition accuracy, but also for representation consistency, transmission robustness, and controllable reconstruction. After selecting and structuring the visual content, the next key issue is whether the compressed, transmitted, rendered, or generated media remains perceptually faithful, temporally coherent, and useful for users or downstream tasks.

	\subsubsection{Media Quality Assessment}

	After content representation, an important question is whether the received or reconstructed media can provide satisfactory perceptual quality for human users and sufficient utility for machine agents. Traditional full-reference metrics, such as PSNR and SSIM, remain useful for conventional compression when pristine references are available and distortion types are predictable. However, these assumptions become less reliable in 6G vision services, where media may be compressed, enhanced, rendered, or generated at the receiver. In such scenarios, users attend to different regions unequally, and visually sharp content may still be inconsistent with the source or unsuitable for the intended task. Therefore, media quality assessment should move beyond signal-level distortion measurement and provide feedback that reflects perception, content importance, user preference, and task utility.

	Deep perceptual quality models provide an important step toward this goal. Feature-space metrics can better capture texture, structure, and visual appearance than pixel-wise errors alone~\cite{ding2020image}. Since pristine references are often unavailable in practical networks, no-reference quality assessment becomes essential for online monitoring at terminals and edge nodes. Hypernetwork-based blind image quality assessment adapts quality prediction to diverse content~\cite{su2020blindly}, while multi-scale Transformer models improve robustness to different resolutions and aspect ratios~\cite{ke2021musiq}. For video services, recent studies further consider high-resolution and user-generated videos by jointly modeling technical distortion and aesthetic perception~\cite{wu2022fast,wu2023exploring}. These works extend quality assessment toward reference-free, perceptually consistent, and low-latency feedback.

	Nevertheless, a single quality score is insufficient for many 6G vision services. Distortions on key objects, interactive regions, or salient areas often affect user experience more strongly than background degradation. Therefore, quality assessment should also consider where distortion occurs and how important the affected content is. Attention- and preference-aware models provide useful mechanisms for content-weighted quality prediction~\cite{talebi2018nima}. More recently, MLLMs have extended quality assessment from score prediction to explainable and comparative quality understanding~\cite{wu2023q,you2024depicting}. By aligning visual quality with human-defined quality levels and interactive reasoning, these models can provide fine-grained feedback, such as whether a key object is blurred, temporal consistency is broken, or a downstream task may be affected~\cite{zhang2024q,ge2025lmm}.

	Generative vision services further broaden the scope of media quality assessment. When receiver-side reconstruction relies on generative models, distortions are no longer limited to compression artifacts, noise, or blur. They may also appear as semantic drift, identity inconsistency, object deformation, temporal flickering, physical implausibility, or text-visual mismatch. Accordingly, recent studies evaluate generated images by considering authenticity, naturalness, preference, and prompt consistency~\cite{li2023agiqa}. For generated videos, benchmarks such as VBench and EvalCrafter examine subject consistency, motion plausibility, temporal stability, and text-video alignment~\cite{huang2024vbench,liu2024evalcrafter}, while other studies further analyze frame-level quality and temporal coherence~\cite{lu2024aigc}. In 6G receiver-side reconstruction, these dimensions determine whether compact cues or conditional information can be transformed into usable visual experience. Therefore, quality assessment should be embedded into the media communication loop to guide bitrate allocation, frame-rate adaptation, edge offloading, bandwidth scheduling, and receiver-side enhancement. However, perceptual quality and task utility alone cannot guarantee that media content is trustworthy, privacy-preserving, or suitable for dissemination, which further motivates content security and compliance detection.
	
	\subsubsection{Content Security and Compliance Detection}

	Media quality assessment focuses on perceptual quality and task utility, but visually high-quality content is not necessarily authentic, privacy-preserving, compliant, or safe. In future media communication systems, visual streams may contain sensitive identities, private scenes, confidential information, manipulation traces, and AI-generated content. Therefore, content security and compliance detection should be integrated into the communication pipeline as a real-time protection mechanism rather than treated as an offline moderation process.

	A primary concern is sensitive information leakage. Faces, license plates, identity documents, medical images, and industrial scenes may expose personal or organizational privacy. Simple blurring or masking can reduce leakage, but may also damage visual quality and task utility. Recent approaches combine detection, tracking, and generative anonymization to preserve useful visual information while protecting privacy~\cite{maximov2020ciagan,hukkelaas2023deepprivacy2}. In 6G systems, such functions are expected to operate at terminals or edge nodes together with anonymization, encryption, access control, and privacy-aware transmission policies.

	Content integrity and authenticity form another challenge. Visual content may be manipulated through splicing, inpainting, replay attacks, or semantic-level editing. Modern image forensics employs deep learning for manipulation detection, localization, and attribution~\cite{mehrjardi2023survey}, while video benchmarks support evaluation under compression artifacts and temporal inconsistencies~\cite{rossler2019faceforensics++,li2020celeb}. For media communication, authenticity assessment should provide manipulation regions, confidence levels, and recommended actions, rather than only a binary real/fake decision.

	The rise of generative AI further increases detection difficulty. Synthetic media produced by diffusion-based generators may be visually realistic and difficult to distinguish after compression, enhancement, or network transmission. Existing studies have explored generator traces, cross-generator generalization, and AIGC attribution~\cite{wang2020cnn,ojha2023towards}. Practical detectors for 6G media services should remain effective under unknown generators, post-processing operations, and real-time constraints, while reporting the risk type, possible source, and confidence level.

	Since passive detection can be affected by data bias, adversarial attacks, and unseen generators, content security is increasingly moving toward proactive verification. Neural watermarking embeds invisible fingerprints into generated content~\cite{fernandez2023stable}, while provenance mechanisms record creation and editing histories. In 6G vision communication, watermarking, provenance tracking, passive detection, and policy-aware moderation should work together to support trustworthy media transmission and generative reconstruction.

	\subsection{AIGC for Media}
	
	\subsubsection{Generative Media Compression}

	Generative Media Compression (GMC) introduces a new paradigm for media coding. Conventional compression systems mainly aim to represent and transmit visual signals as efficiently as possible for accurate reconstruction. In contrast, GMC exploits generative models to recover visual details that are not explicitly transmitted. Instead of delivering complete pixel-level information, the transmitter only sends compact conditions, such as structural cues, motion descriptors, latent representations, or low-resolution references. The receiver then reconstructs perceptually plausible media content with the help of generative priors. In this sense, the role of communication gradually shifts from direct signal delivery to generation-guided reconstruction.

	Early studies explored the use of generative priors in image and video compression. GAN-based image compression methods showed that visually plausible textures can be recovered from highly compressed representations by learning natural image distributions. Similar ideas were later extended to generative facial video coding~\cite{Siarohin2019fomm,chen2022beyond}. In these systems, the transmitter sends sparse motion or keypoint descriptors, while the receiver synthesizes realistic facial frames using a generative model. Although such methods are often designed for human-centric scenarios, they establish an important principle of generative compression: only compact guidance information needs to be transmitted, while missing visual details can be reconstructed by the receiver-side generative model.

	To extend generative coding beyond specific domains, recent studies have investigated learned latent representations for general visual content. Representative frameworks~\cite{guo2025glvc,zheng2026gvc1d} encode visual content into compact latent spaces and employ generative decoders to reconstruct details. Compared with conventional learned compression methods that mainly pursue deterministic signal reconstruction, these approaches place greater emphasis on perceptual reconstruction. By leveraging learned generative priors, they can produce visually convincing results under ultra-low bitrate conditions, where transmitting all fine-grained textures is usually infeasible.

	More recently, diffusion models have become powerful generative priors for media compression. Unlike GAN-based reconstruction, diffusion models recover visual content through progressive denoising, which provides strong distribution modeling capability and stable reconstruction quality. Representative approaches~\cite{xue2025s2vc,wang2026disco} demonstrate that high-quality media content can be reconstructed from sparse conditions, low-resolution references, or structural constraints. This direction further reduces the amount of information that must be explicitly transmitted, while maintaining perceptual consistency and reconstruction controllability. Therefore, diffusion-based compression provides a promising solution for bandwidth-constrained 6G media services, especially when the objective is to preserve perceptual quality rather than exact pixel-level fidelity.

	Looking forward, the evolution of GMC is increasingly influenced by foundation models and tokenized media representations. Techniques such as VQ-VAE~\cite{van2017vqvae} and VQ-GAN~\cite{esser2021taming} represent visual content as compact discrete tokens that can be compressed, transmitted, and interpreted by generative decoders. As MLLMs and VLMs improve visual understanding and generation, GMC can jointly optimize compression, reconstruction, and generation without repeatedly transmitting full visual details.

	\subsubsection{Personalized Context Adaptation}

	While generative media compression changes how media content is represented and transmitted, personalized context adaptation (PCA) further focuses on how content is delivered, reconstructed, and experienced by different users. Conventional communication systems often follow a ``one-content-for-all'' paradigm, where the same media stream is delivered with limited consideration of device capability, user preference, environmental context, or interaction state. As media services become increasingly interactive and user-specific, this paradigm is no longer sufficient. Therefore, media communication is evolving from universal content delivery toward personalized experience generation.

	Early personalized adaptation mainly focused on context-aware delivery. In this stage, media quality is adjusted according to network conditions, device capabilities, and service requirements. Representative studies combine adaptive streaming, AI-enhanced reconstruction, and efficient generative models to improve the balance between communication efficiency and quality of experience (QoE) under dynamic transmission environments~\cite{lu2025abuv}. However, these methods mainly optimize the delivery of existing content, while the content itself is usually not deeply adapted to users or surrounding contexts.

	With the development of generative models, PCA is further extended to environment-aware and preference-aware generation. Instead of only adjusting bitrate, resolution, or frame rate, generative models can modify viewpoints, scene details, and rendering fidelity according to contextual information and user interaction. Representative frameworks show that immersive scenes can be generated from compact conditions without transmitting complete scene representations~\cite{zhou2024dreamscene360}. In addition, preference alignment techniques enable generated media to better match users' interests, viewing habits, and subjective expectations~\cite{rafailov2024dpo,liu2025videodpo}. Under this paradigm, personalization is no longer limited to delivery adaptation, but extends to the generated content itself.

	More recently, multi-modal foundation models and interactive AI systems have enabled continuous personalization through real-time interaction. Advanced frameworks integrate content understanding, user intent modeling, and media generation within a unified architecture~\cite{zhang2025videollama3,xiong2025streamchat}. In such systems, user feedback, dialogue, gaze, gestures, and interaction history can serve as contextual cues for media generation and refinement. As a result, personalization becomes an interactive co-creation process, where media content is dynamically updated according to both user behavior and communication conditions.

	PCA thus moves media delivery from context-aware streaming toward environment-aware, preference-aware, and interactive generation. Together with generative media compression, it allows context, user intent, and preference-aware generation to become explicit control variables for personalized 6G media services.

	\subsubsection{Virtual Content Generation}

	While generative media compression focuses on compact representation and reconstruction, virtual content generation further expands the communication object from video frames to renderable, controllable, and synchronizable virtual scene states. Future media communication may need to represent not only visual appearance, but also scene geometry, object motion, user poses, interaction states, and generative conditions. In this sense, virtual content generation transforms sparse observations, semantic descriptions, and control signals into virtual scenes that can be rendered, updated, and shared across users.

	A key foundation is renderable 3D scene representation. Neural radiance fields (NeRFs) enable novel-view synthesis from multi-view images, while efficient variants reduce representation and rendering costs~\cite{muller2022instant}. More recently, 3D Gaussian Splatting achieves high-quality real-time rendering ~\cite{kerbl20233d}, and dynamic extensions further model time-varying scenes~\cite{wu20244d}. These advances suggest that communication systems may transmit sparse views, camera poses, scene primitives, or compact 3D representations instead of dense visual streams.

	Beyond scene representation, virtual content generation also requires efficient 3D asset creation from text, images, or sparse observations. Diffusion-based methods generate 3D content from language conditions~\cite{lin2023magic3d} and improve generation quality through advanced optimization techniques~\cite{wang2023prolificdreamer}. Multi-view generation and feed-forward reconstruction enable rapid 3D asset creation from limited inputs~\cite{long2024wonder3d}. Gaussian representations and structured 3D latents further improve efficient generation~\cite{xiang2025structured}. These methods make it possible to generate or update virtual objects and scenes from compact conditions, reducing the need to repeatedly transmit dense visual data.

	Human-centric virtual content provides another important direction. Instead of continuously transmitting full facial or body video, compact signals such as speech, expression, pose, gaze, and motion can be delivered, while realistic avatars or human representations are rendered near the receiver \cite{qian2024gaussianavatars}. Video diffusion models further extend generation to temporally coherent clips, supporting image-to-video, animation, and controllable video synthesis~\cite{blattmann2023stable,chen2024videocrafter2}. Virtual content generation therefore shifts media transmission from frame-level pixels to scene-level conditions, such as prompts, poses, motion parameters, Gaussian primitives, and 3D latents. Practical deployment still requires real-time inference, cross-user consistency, identity preservation, physical plausibility, copyright protection, and safety compliance for reliable virtual media communication in 6G networks.
	
	\begin{figure*}[htbp]
	\centering
	\includegraphics[width=6.7in]{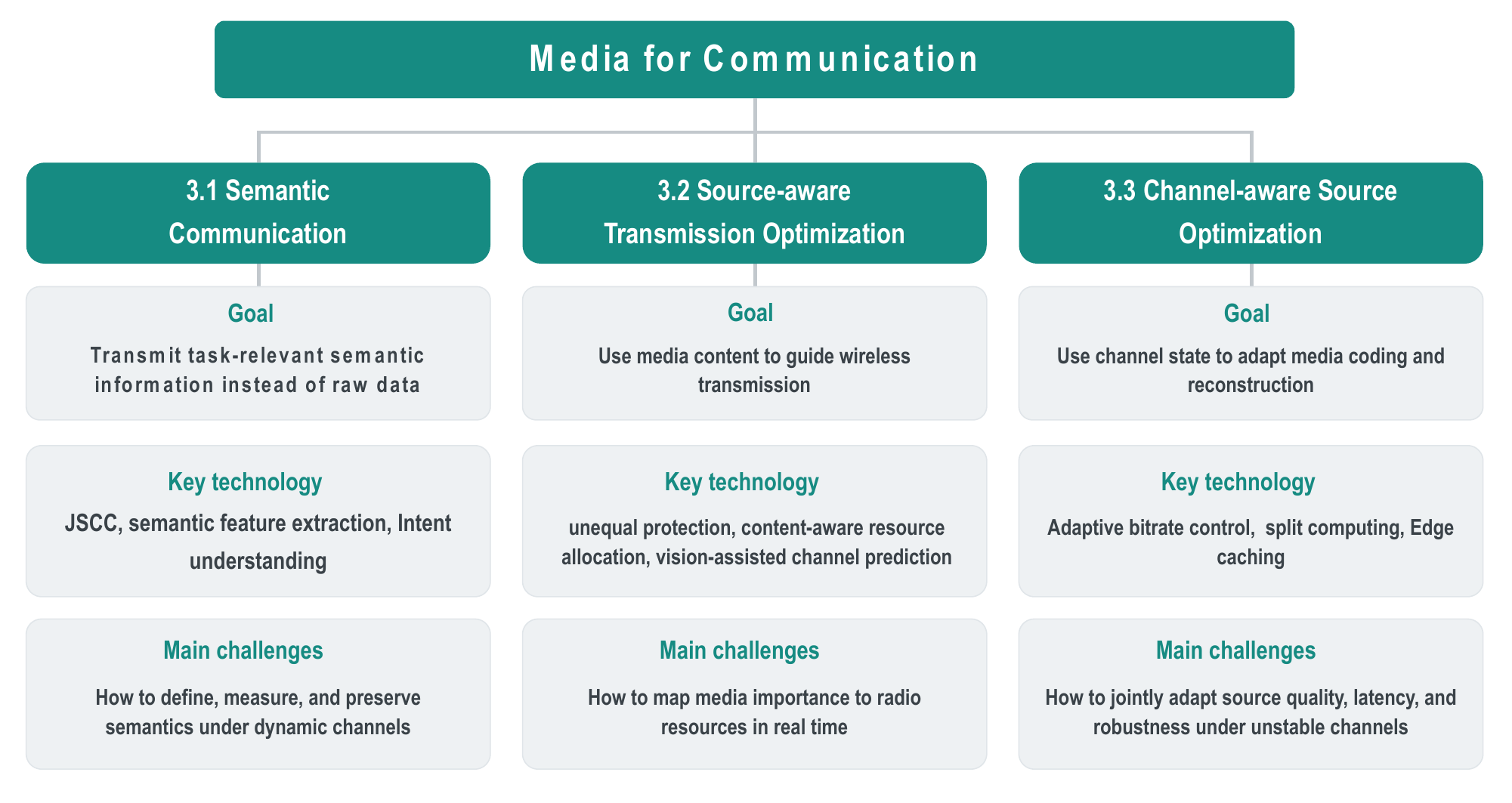}
	\caption{Overview of media-aware wireless transmission.}
	\label{fig_5}
	\end{figure*}	

	\section{Media for Communication}
	In this section, we investigate how media content can become an active design variable in wireless communication systems. Different from conventional pipelines that separately optimize source coding and channel transmission, media-oriented communication exploits semantic structures, content statistics, and task priors to improve transmission efficiency and robustness. Specifically, we first review semantic communication, where joint source-channel coding (JSCC) is used to transmit task-relevant semantic information rather than raw data. Then, we discuss source-aware transmission optimization, in which media characteristics guide beamforming, power allocation, and modulation-scheme selection. Finally, we present channel-aware source optimization, where real-time channel conditions reshape media encoding, edge-assisted processing, and receiver-side reconstruction. These paradigms form a coupling spectrum from fully joint design to one-way source-to-channel and channel-to-source adaptation. The architecture of this section is given in Fig. \ref{fig_5}.

	\subsection{Semantic Communication: Joint Source-Channel Optimization}
	Recently, semantic communication has emerged as a promising paradigm for media transmission \cite{gunduz2023beyond}. The key idea is to prioritize important and task-relevant semantic information instead of transmitting all raw data. By incorporating JSCC techniques, semantic communication can integrate media-specific structures, such as visual saliency, task relevance, and perceptual priors, into the transmission process, thereby enabling more efficient and robust delivery under challenging channel conditions. This shift is particularly important for bandwidth-limited and latency-sensitive media services, where raw data usually contain substantial redundancy.

	In the following, we review recent studies on media semantic communications from three aspects. First, semantic-channel joint optimization focuses on the co-design of semantic feature extraction, channel encoding, and robust reconstruction under noisy wireless channels. Second, we discuss intent-aware transmission scheduling, where the transmitter prioritizes semantic components based on service intent, receiver requirements, and task objectives. Finally, we present recent efforts on cross-layer optimized frameworks, where semantic communication is integrated with higher-layer protocols and network resource management to achieve end-to-end performance optimization.
	
	\subsubsection{Semantic-Aware Joint Source-Channel Coding}
	Semantic-aware JSCC provides one of the most direct implementations of semantic communication for wireless media transmission~\cite{xu2023commmag,CL_SemCom}. Different from the conventional separated source and channel coding architecture, JSCC directly learns the mapping from source samples to channel symbols under distortion-, perception-, or task-aware objectives. Therefore, the transmitted representation can be jointly optimized for channel robustness, semantic fidelity, and receiver-side utility. Early neural JSCC studies mainly focused on reconstruction-oriented point-to-point transmission. DeepJSCC~\cite{sec4-9-bourtsoulatze2019deep} maps images directly to channel symbols and avoids the cliff effect of conventional digital transmission under low-SNR conditions. Subsequent studies introduced feedback and progressive refinement~\cite{kurka2020deepjsccf}, as well as bandwidth-agile transmission with successive refinement and multiple descriptions~\cite{kurka2021bandwidth}, enabling reconstruction quality to adapt to available channel resources.

	Following this foundation, many studies have improved JSCC robustness under dynamic wireless environments and practical physical-layer constraints. For varying SNRs, attention modules and SNR-adaptive decoding have been used to adapt latent features or decoders to channel states~\cite{xu2022attention}. To bridge the gap between idealized AWGN training and practical broadband transmission, DeepJSCC with OFDM embeds differentiable OFDM processing and frequency-selective multipath channels into the neural transceiver~\cite{yang2021ofdm}. DeepJSCC-l++ further supports multiple bandwidth ratios and SNR levels within a single model~\cite{bian2023deepjsccLpp}. Beyond single-antenna links, JSCC has been extended to MIMO scenarios through learnable CSI fusion~\cite{xie2024lcfsc}, self-attention and power allocation~\cite{wu2024mimo}, multi-user MIMO access with cooperative successive interference cancellation~\cite{xie2025multiuserMimo}, and variable-length and variable-rate video coding for MIMO services~\cite{xie2026cvst,xie2026contextual}.

	In addition to robustness-oriented design, semantic-aware JSCC architectures have evolved from convolutional transceivers to attention- and Transformer-based models. Early neural JSCC systems mainly relied on CNNs, which are effective for local feature extraction but limited in capturing long-range dependencies and global semantic structures. Attention-based feature modulation improves representation adaptability by adjusting latent features according to channel conditions~\cite{xu2022attention}. Building on this idea, Transformer-based JSCC architectures provide stronger global modeling and flexible transmission. DeepJSCC-l++ adopts a unified Transformer-based model for multiple bandwidth ratios and SNRs~\cite{bian2023deepjsccLpp}; WITT redesigns the vision Transformer for wireless image transmission with channel-adaptive feature modulation~\cite{yang2023witt}; and SwinJSCC incorporates Swin Transformer blocks to improve multi-scale representation across resolutions, rates, and channel states~\cite{yang2025swinjscc}. These studies show that Transformers enhance semantic modeling, channel adaptation, and feature-level resource allocation in JSCC systems.

\begin{figure*}[htbp]
	\centering
	\includegraphics[width=6.9in]{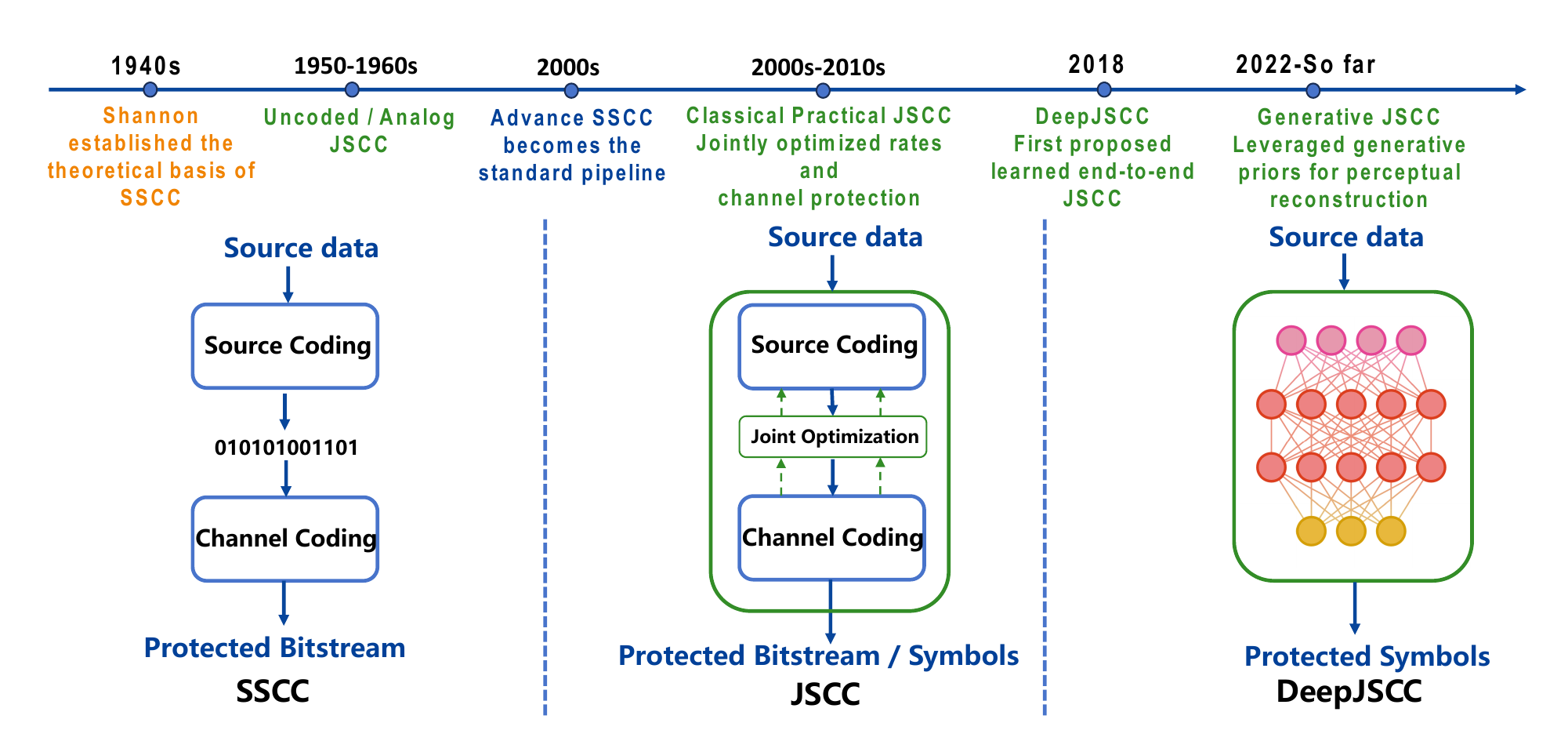}
	\caption{Evolutions from SSCC to JSCC.}
	\label{fig_6}
	\end{figure*}

	More recently, generative models have opened a new direction for semantic-aware JSCC by transmitting compact semantic conditions and exploiting receiver-side priors for reconstruction, completion, or task inference \cite{RAG_Semcom}. GAN-based generative JSCC improves visual quality under severe bandwidth or channel constraints through adversarial perceptual priors~\cite{erdemir2023generative}, while channel-aware GAN inversion extracts semantic information that can be directly transmitted and reconstructed by a receiver-side generator without retraining for different channels~\cite{tang2024evolving}. Diffusion models further strengthen this paradigm with more powerful generative priors. DiffJSCC combines Stable Diffusion with multi-modal visual features and CSI to improve realism in low-bandwidth and low-SNR regimes~\cite{rombach2022latentDiffusion,yang2025diffjscc}. Semantics-guided diffusion JSCC uses text or edge-map conditions for visual recovery~\cite{zhang2026sgdJscc}, while training-free diffusion-based semantic communication reduces task-specific retraining through diffusion inversion and sampling~\cite{tang2025trainingFreeDiffusion}. Related extensions further exploit reusable semantic codebooks for cache-enabled transmission~\cite{tang2026cacheEnabled} and diffusion-aided decoding for secure and robust JSCC against attacks such as subcarrier jamming and pilot spoofing~\cite{zhao2026secdiff}. These works indicate that generative semantic JSCC shifts transmission from delivering complete source details to delivering sufficient semantic guidance for receiver-side generation, robust reconstruction, and task inference. As shown in Fig. \ref{fig_6}, the coupling between source coding and channel coding has gradually evolved from SSCC to different forms of JSCC.
	
	\subsubsection{Intent-aware Transmission Scheduling}
	Semantic-aware JSCC mainly determines how semantic representations are encoded, whereas intent-aware transmission scheduling determines which semantic components should be transmitted, updated, retransmitted, or discarded. In media services, visual components have different values for the receiver's goal. Thus, the scheduling unit is no longer only a homogeneous packet or pixel block, but a semantic unit whose priority depends on the task, user intent, and decision context. This idea is reflected in goal-oriented communication, which shifts the communication objective from signal fidelity to goal accomplishment~\cite{zhang2022goalIot}. Feature-pruning studies further show that redundant visual features can be removed for edge inference~\cite{shao2023taskEdgeInference}, while task-oriented ARQ indicates that retransmission can be triggered by inference confidence and CSI rather than only by bit errors \cite{li2024tarq}.
	
	Based on this principle, existing studies generate priority labels from visual saliency, region of interest (ROI), semantic importance, and user context. Classical ROI- and saliency-aware schemes provide early examples of unequal visual protection, where important regions are assigned stronger coding or transmission resources \cite{huo2019roiUep}. Recent semantic communication studies further move from perceptual saliency to task-dependent semantic value. Personalized saliency and scene semantics can be used to select visual information according to user interests~\cite{kang2023personalizedSaliency}. Semantic-importance transmission \cite{sun2023semanticsPixels} assigns different weights to features or pixels according to their contribution to downstream perception. User-context-aware schemes further use gaze, operation history, and inferred intention to select geometry, texture, object pose, text, layout, or scene semantics according to service goals \cite{wang2025metaverseGsc,jiang2026aiGlasses}. These studies show that intent-aware scheduling can operate at different granularities, including objects, regions, patches, feature channels, and video segments.
	
	Moreover, intent-aware scheduling should also consider temporal freshness. A semantic update may quickly lose value after the object moves, the gaze point changes, or the inference decision has already been made. Value of information (VoI) and age of incorrect information (AoII) provide useful tools for modeling whether a semantic update is still worth transmitting \cite{ayan2019aoiVoi,maatouk2023aoii}. AoII-based scheduling applies information freshness to transmission decisions by jointly considering semantic error and information age~\cite{chen2023aoiiXr}. Freshness-aware DeepJSCC \cite{basnayaka2024aomiDjscc} and timeliness-aware adaptive JSCC \cite{yang2025timelinessJscc} further optimize transmission according to classification correctness, reconstruction quality, and VoI. Intent-aware scheduling therefore complements semantic-aware JSCC by deciding what should be communicated and when it remains valuable, leaving the mapping from semantic priority to wireless resources to cross-layer optimization.

	\subsubsection{Cross-Layer Optimized Framework}
	To fully exploit semantic communication, semantic-channel JSCC and intent-aware scheduling should be further connected with wireless resource management. Specifically, source semantics, CSI, queueing states, and radio resources need to be jointly optimized, rather than separately designing the JSCC transceiver, application-layer priority map, and radio scheduler. In this way, semantic-priority information can guide resource block (RB), subcarrier, bandwidth, and power allocation according to reconstruction quality, task accuracy, or QoE \cite{liu2024ofdmImportance,liu2025siareOfdm}.
	
	Existing studies have mainly formulated this problem as semantic-aware resource allocation and joint compression--transmission optimization. Several studies incorporate semantic similarity and QoE into channel assignment, bandwidth allocation, and power control for semantic transmission~\cite{yan2022textResource,yan2022qoeResource}. Another line of work jointly optimizes the semantic compression ratio and wireless resources~\cite{liu2024ascResource}, while feature-importance-aware task-oriented transmission couples feature selection with channel assignment through reinforcement learning (RL)~\cite{wang2024featureImportance}. Moreover, a DRL-based task-oriented semantic communication framework optimizes the compression ratio, transmit power, and bandwidth according to task accuracy, showing that resource allocation can be driven by task utility rather than bit-level reliability alone \cite{zhang2023drlTaskResource}. This cross-layer view has also been extended to practical service and channel models. In vehicular, device-to-device (D2D), and multi-task networks, dynamic allocation schemes jointly consider semantic priority, channel quality, and heterogeneous service requirements in the scheduler \cite{yan2024qoeMultitask,wang2024adaptiveResource}. For media-oriented JSCC and digital semantic systems, channel uses, latency, and capacity can be optimized under distortion or quality of service (QoS) constraints \cite{chi2025capacityQos}. OFDM-based designs \cite{liu2024ofdmImportance,liu2025siareOfdm} provide a concrete PHY-layer realization by mapping important semantic features to more reliable subcarriers or assigning more bits and power to them. In multi-device video analytics, a representative framework jointly coordinates sensing selection, semantic extraction, and wireless transmission according to downstream analytics objectives~\cite{he2025sensem}. Furthermore, queueing-aware semantic communication treats the latent dimension as a controllable transmission variable, thereby incorporating service time, queueing delay, and information freshness into semantic communication \cite{sagduyu2026queueing}.
	
	These studies also motivate new cross-layer key performance indicators (KPIs). Beyond bit rate, block error rate, and packet delay, semantic throughput can measure the amount of task-useful or perceptually valid information delivered per unit time or spectrum \cite{yan2022textResource,wang2024featureImportance}. Semantic outage further describes the event that the received semantics cannot satisfy the required reconstruction quality, inference accuracy, or interaction reliability \cite{liang2025siacMimo}. Specifically, the application layer should provide compact semantic-priority labels, the MAC layer should translate labels into scheduling, retransmission, and queue-management decisions, and the PHY layer should map them to modulation, coding, power, and subcarrier choices. A protocol-level design further shows that lower layers can preserve noisy semantic payloads for application-layer recovery, instead of always enforcing bit-perfect retransmission~\cite{wang2025sitp}.
	
	\subsection{Source-aware Transmission Optimization}
	In addition to semantic-aware JSCC, media content can also guide the design of conventional transmission schemes. Different from end-to-end semantic-aware JSCC, source-aware transmission optimization exploits the characteristics of visual sources to improve efficiency and robustness of existing wireless mechanisms, such as modulation, channel coding, power allocation, and scheduling. Moreover, visual observations can also provide environment knowledge for channel prediction and compensation. This subsection reviews source-aware transmission optimization from two perspectives: media content-aware channel adaptation and vision-assisted channel prediction and compensation. 

	\subsubsection{Media Content-aware Channel Adaptation}
	In wireless media delivery, the loss of different visual units has different effects on the final experience. For example, losing an I-frame, a reference frame, or a scalable video coding (SVC) base layer may affect a group of dependent frames, while losing a background texture block may only introduce local degradation. Similarly, in panoramic and immersive video, the loss of an ROI or viewport-centered tile is usually more harmful than the loss of a peripheral tile. Therefore, it is inefficient to assign the same modulation and coding scheme (MCS), packet priority, or transmit power to all packets. The key problem is how to expose such visual importance to the transmission layer and then map it to unequal protection and wireless resource allocation.

	Existing studies usually represent visual importance through distortion cost, dependency structure, saliency, or delay deadline. For example, the content-aware resource allocation framework in \cite{pahalawatta2007contentAwareResource} considers pre-encoded video packets in wireless downlink systems, and uses packet-level distortion impact to guide resource allocation and scheduling. The perceptually unequal loss protection method in \cite{ha2010pulp} jointly considers visual saliency and error propagation, so that packets affecting salient regions or later frames in the GoP receive stronger protection. Moreover, the channel, deadline, and distortion scheduler in \cite{dua2010cd2} assigns transmission opportunities according to channel quality, packet deadline, and packet distortion cost. These studies show that the transmission layer can operate on compact importance indicators, instead of treating all packets as homogeneous bit payloads.

	Based on such indicators, one representative direction is unequal protection and priority scheduling. In \cite{ha2010pulp}, perceptual importance and temporal error propagation are translated into forward error correction allocation, thereby reducing perceptual degradation under packet-erasure channels. In \cite{pahalawatta2007contentAwareResource}, packet scheduling and wireless resource allocation are jointly optimized for video transmission over wireless networks, where packets with larger distortion impact are given higher scheduling priority. The content-aware distortion-fair scheme in \cite{li2009distortionFair} further exploits temporal prediction structure and frame-drop distortion to decide which video frames should be dropped under congestion, so that the system improves distortion fairness among competing video streams. These works represent an application-to-MAC interface: visual importance is measured at the media layer, while packet scheduling and dropping are performed at the transmission layer.

	Another direction is to incorporate visual importance into PHY-layer and radio-resource decisions. For layered video, the base layer is more important because enhancement layers depend on it. Therefore, cross-layer SVC transmission maps the base layer to more reliable channel configurations, while enhancement layers can use higher-rate configurations when the channel condition permits \cite{zhang2010crossLayerSvc}. Distributed media-aware rate allocation further exploits rate-distortion characteristics to allocate wireless bandwidth among multiple video streams, thereby balancing the quality of competing sessions \cite{zhu2010mediaAwareRate}. In addition, significance-aware power allocation assigns more power to important video frames \cite{hong2010significancePower}, while distortion-aware power allocation in multi-user non-orthogonal multiple access (NOMA) systems embeds video distortion models into power control \cite{lu2021distortionNoma}. These studies show that media content can guide not only packet priorities, but also MCS selection, subcarrier assignment, bandwidth allocation, and power control.

	\subsubsection{Vision-assisted Channel Prediction and Compensation}
	The aforementioned content-aware schemes mainly exploit visual information after the media packets or features have been generated. In contrast, vision-assisted channel prediction uses visual sensing to infer the wireless environment before severe channel degradation is observed from RF measurements. This direction is particularly important for mmWave, sub-THz, and THz systems, where narrow beams, mobility, human blockage, and sparse scattering make beam training and CSI acquisition costly. Specifically, cameras, depth sensors, LiDAR, radar, and position sensors can provide out-of-band information about user locations, obstacle trajectories, line-of-sight (LoS) states, and surrounding scatterers. Therefore, the visual scene can be regarded as a physical prior for channel prediction and compensation, instead of only being treated as the media source to be transmitted. Early RGB-D and depth-image-based studies have shown that human blockage and received-power degradation can be predicted before the mmWave link collapses \cite{oguma2016rgbdHandover,nishio2019receivedPower}. Moreover, DeepSense 6G further provides a large-scale real-world benchmark with synchronized wireless, camera, GPS, LiDAR, and radar measurements, which has promoted the evaluation of vision-assisted beam, blockage, and positioning algorithms \cite{alkhateeb2023deepsense6g}.

	Existing works first investigate vision-assisted beam management and blockage compensation. The camera-aided mmWave base-station framework in \cite{alrabeiah2020cameraBeam} predicts the best beam and blockage state from RGB images, thereby reducing the overhead caused by exhaustive beam sweeping. Vision-aided blockage prediction further enables proactive beam switching or handoff before LoS blockage causes link outage \cite{charan2022visionBlockage}. For mobile users, vision-aided beam tracking exploits historical image sequences to predict future beams, while prototype-level multi-user beam tracking demonstrates the feasibility of extending this idea to mmWave massive MIMO systems with multiple moving users \cite{li2024vamubt}. In addition to camera-only solutions, LiDAR-aided and radar-aided beam prediction use point clouds or range-angle maps to improve robustness under low-light conditions, adverse weather, or visual occlusion \cite{demirhan2022radarBeam}. Furthermore, position-aided beam prediction and multi-modal Transformer models jointly exploit camera, GPS, radar, LiDAR, and temporal information, which shows that different sensing modalities should be fused according to their reliability in practical deployment \cite{tian2023multimodalTransformer}.

	Beyond discrete beam and blockage labels, recent studies have extended vision-assisted prediction to more general channel-state inference and compensation. Depth image-based learning predicts future received-power variations, while RGB-image- and segmentation-based methods estimate received power, path loss, Rician $K$-factor, and RMS delay spread in vehicular or street-intersection scenarios \cite{nishio2019receivedPower,zhang2025visionPower}. In WiFi sensing, visual information has also been used to compensate for incomplete or compressed RF observations by recomposing CSI from beamforming feedback matrices \cite{shimomura2023visionCsi}. From the channel-map perspective, CKMImageNet pairs visual representations with channel knowledge maps, enabling computer vision models to learn location-specific channel priors for environment-aware channel estimation and resource allocation \cite{wu2024ckmimagenet}. These studies indicate a general workflow: the perception module first extracts geometry and motion-related channel priors from the visual scene, and the transmission module then adjusts beams, handoff decisions, pilot overhead, or CSI recovery before the channel degradation becomes severe.

	Another promising direction is to incorporate visual perception into geometry-based and digital-twin channel modeling. Digital twin channels construct a virtual representation of the propagation environment by combining online measurements, environmental maps, and ray-tracing models, thereby supporting proactive network control \cite{wang2025dtc}. Differentiable ray tracing further enables the material and propagation parameters of the radio environment to be learned from measurements, which mitigates the mismatch between simulated and deployed channels \cite{hoydis2024differentiableRt}. More recently, large model- and VLM-based methods have been introduced to enhance this pipeline. Specifically, ChannelGPT explores large model-based generation of digital-twin channel parameters, VLM-guided beam prediction aligns images, LiDAR, and location prompts for cross-modal representation learning, and VLM-guided differentiable ray tracing uses semantic material priors to accelerate RF parameter estimation \cite{yu2024channelgpt,wang2026vlmBeam,kang2026vlmDrt}. These works highlight that visual semantics can serve as an interface between scene understanding and physics-based channel inference.

	\begin{figure*}[htbp]
	\centering
	\includegraphics[width=6.6in]{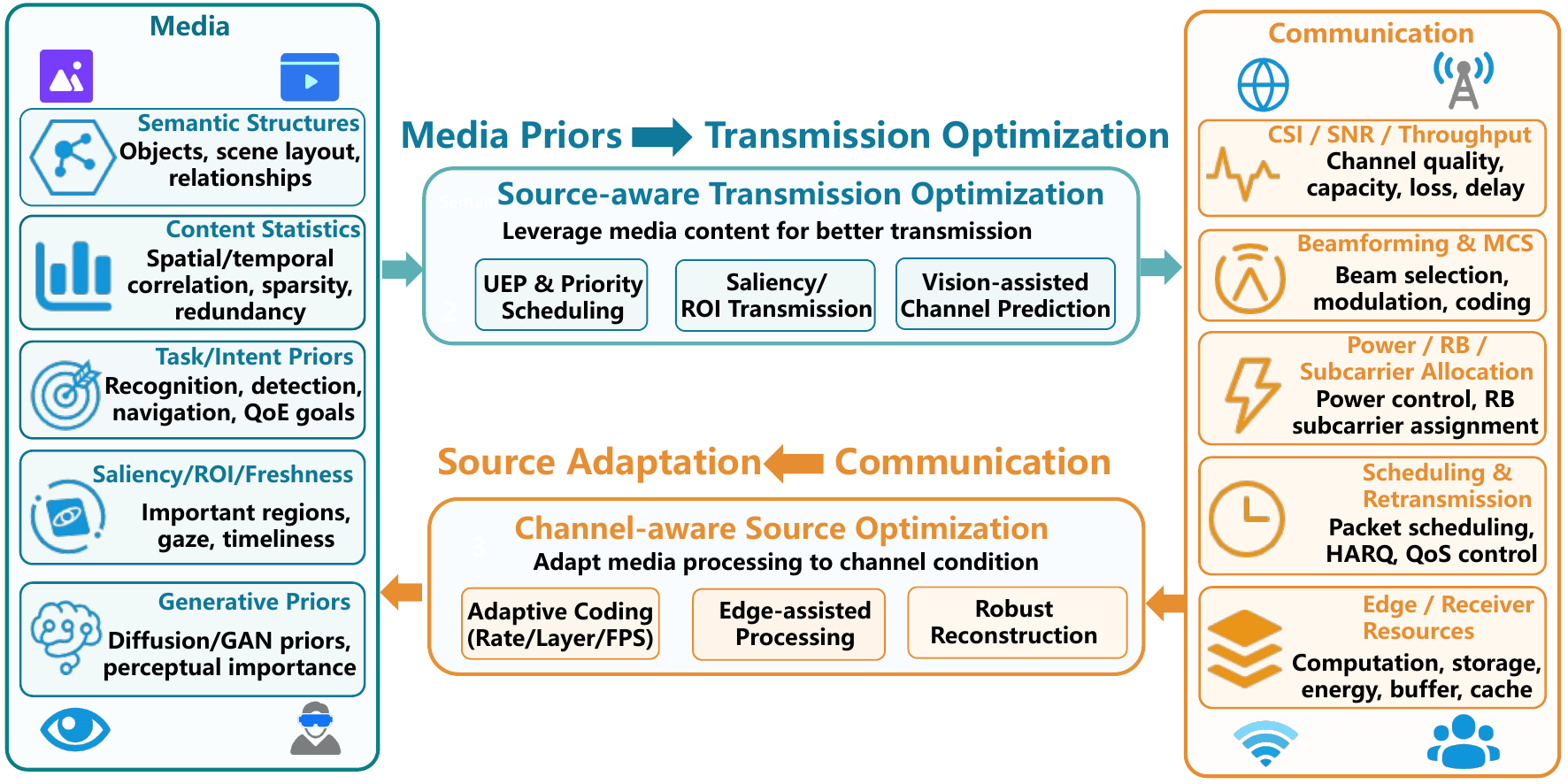}
	\caption{Relationship between channel-aware and source-aware operations.}
	\label{fig_7}
	\end{figure*}

	\subsection{Channel-aware Source Optimization}

	In addition to source-aware transmission, channel state can also guide source-side media coding and processing. This paradigm reverses the previous direction: instead of using media features to assist the wireless link, the transmitter, edge node, or receiver uses real-time channel feedback to reshape the visual source representation. Such feedback may include throughput, SNR, packet loss, buffer occupancy, latency budget, and lower-layer scheduling information, enabling the source pipeline to adjust bitrate, resolution, frame rate, quality layer, feature dimension, offloading point, or reconstruction strategy before severe quality degradation occurs. Accordingly, this subsection reviews channel-aware source optimization along the data path, including transmitter-side adaptive media coding, edge-enhanced collaborative processing, and receiver-side robust reconstruction under impaired channels. Moreover, the relationship between channel-aware and source-aware operations is shown in Fig. \ref{fig_7}.

	\subsubsection{Channel-aware Adaptive Media Coding}
	The most direct form of channel-aware source optimization is adaptive media coding, where the source encoder or streaming controller changes its output according to the observed network state. In conventional adaptive bitrate (ABR) streaming, the channel is usually observed through application-level proxies such as measured throughput and playback buffer occupancy. Buffer-based adaptation showed that buffer occupancy alone can serve as an effective signal for selecting the next video rate \cite{huang2014bba}, while control-theoretic ABR further predicts future bandwidth and optimizes the bitrate sequence under buffer constraints \cite{yin2015mpc}. Learning-based approaches replace manually designed control rules with policies trained to optimize QoE. Pensieve uses RL to select video bitrates under changing throughput and buffer states \cite{mao2017pensieve}, whereas Oboe tunes ABR parameters online according to the current network condition \cite{akhtar2018oboe}. More recent controllers incorporate perceptual quality and smoothness more explicitly, such as quality-aware imitation learning in Comyco \cite{huang2019comyco} and smoothness-optimized dynamic adaptation in SODA \cite{neufeld2024soda}. These works indicate that even when the codec itself is unchanged, the delivered source representation can be continuously reshaped by channel feedback.

	For low-latency and wireless services, the feedback loop needs to be tighter than segment-level ABR. Salsify integrates the video codec with the transport protocol, allowing the sender to explore multiple compression options for each frame and transmit the version that matches the latest capacity estimate \cite{fouladi2018salsify}. In wireless streaming, exposing lower-layer information to the application controller further improves adaptation because throughput alone may hide fast variations in scheduling, modulation, or link quality \cite{huang2025neuralwireless}. This cross-layer exposure turns channel-aware source coding from a purely application-layer decision into a closed-loop interaction between the media encoder and the wireless stack.

	Layered and scalable representations provide another natural mechanism for channel-aware coding. The scalable extension of H.264/AVC separates video into temporal, spatial, or quality layers, so that a receiver can preserve the base layer under poor channels and opportunistically receive enhancement layers when channel capacity improves \cite{schwarz2007svc}. This structure also supports unequal protection, where the base layer receives stronger forward error correction (FEC), power, or network-coding protection than enhancement layers \cite{shih2010shvcuep,thomos2018rlnc5g}. Neural layered codecs extend the same idea to learned representations. Swift encodes video into neural layers that can be combined at the receiver for fine-grained quality adaptation \cite{dasari2022swift}. Therefore, channel feedback can be mapped not only to a scalar bitrate, but also to a source-layer selection policy.

	\subsubsection{Edge Node-enhanced Collaborative Processing}
	When the source device cannot adapt sufficiently or when multiple users share a constrained wireless link, a nearby edge node can participate in source optimization. The role of the edge in this subsection is not general-purpose computation offloading, but channel-aware media pre-processing before visual data crosses the bottleneck link. In adaptive video delivery, multi-access edge computing (MEC) servers can cache popular segments, transcode them into channel-suitable representations, and coordinate wireless resource allocation. Joint caching, transcoding, and resource allocation frameworks show that edge servers can prepare user-specific video versions according to wireless conditions and QoE requirements \cite{liu2021edgetranscode,tran2019edgeenergy}. Software-defined mobile networks further enable deep reinforcement learning to jointly manage edge caching and transcoding decisions over time-varying wireless links \cite{wang2019drlmec}. In this line of work, the edge converts channel feedback into representation management: which segment should be cached, which bitrate or resolution should be produced, and which user should receive which version.

	Edge collaboration is also important for video analytics, where transmitting full-resolution streams to the cloud or edge server may be unnecessary. Instead of sending every pixel, the front-end device and edge server can jointly select the frame rate, resolution, region of interest, and model configuration. Joint configuration adaptation and bandwidth allocation optimize these parameters under bandwidth, accuracy, energy, and latency constraints \cite{zhang2022tnetjcab}. More recent spatial-temporal semantic filtering systems use tracking at the device and detection at the edge to avoid uploading redundant frames, while dynamically adjusting offloaded frame resolution and model configuration according to the channel and processing deadline \cite{yang2024taodm}. These methods reveal a key difference between media streaming and vision analytics: under poor channels, the best source representation may be a sparse task-oriented stream rather than a lower-quality full video.

	A related direction is split computing or encoding, where early layers of a neural model run on the device and intermediate features are sent to the edge. Neurosurgeon provided an early system-level example of choosing the split point according to latency, bandwidth, and energy tradeoffs \cite{kang2017neurosurgeon}. Later systems show that split points, bottleneck features, and early exits should be adapted to network and edge-load variations rather than fixed offline \cite{matsubara2022survey}. Supervised compression explicitly learns compact intermediate features for resource-constrained edge systems, avoiding the transmission of task-irrelevant visual information \cite{matsubara2022supervised}. Slimmable encoders make this idea more flexible by changing the feature dimension according to the available bandwidth \cite{matsubara2023slimmable}. Progressive neural compression further orders transmitted information by its contribution to inference performance, so the sender can stop transmission when timing or bandwidth limits are reached \cite{chen2023progneural}. Multi-branch and compression-aware split systems extend this adaptation by selecting among multiple compressed branches or compression levels at runtime \cite{navisplit2024}.

	These edge-assisted designs reduce the dependence on raw bandwidth by moving source-side decisions closer to the wireless bottleneck. However, they also introduce new coordination challenges. Edge pre-processing consumes computation and energy, may reveal sensitive visual features, and must remain stable under fluctuating channels and workloads. In distributed settings, edge nodes may need to collaborate on workload placement, model selection, and request routing, as studied in multi-agent edge video analytics frameworks \cite{xie2024edgevision}. Therefore, the central design problem is not simply whether to use the edge, but how to decide which visual information should be transformed, discarded, compressed, or forwarded when channel and edge resources change together \cite{wang2023splitsurvey}.

	\subsubsection{Robust Vision Reconstruction under Poor Channels}
	Even with transmitter-side adaptation and edge-side pre-processing, packet loss, bit errors, burst erasures, and deep fades can still reach the receiver. Robust vision reconstruction addresses this residual impairment by converting channel damage into recoverable visual missingness. Traditional error control and concealment methods already established the basic toolbox, including retransmission, FEC, motion-compensated temporal concealment, spatial interpolation, and hybrid schemes \cite{wang1998errorreview}. The recent shift is to replace hand-crafted concealment rules with learned visual priors, while still respecting the real-time and fidelity constraints of communication systems.

	One line of work integrates loss robustness directly into the video codec or decoder. Swin-VEC uses a Video Swin Transformer-based GAN for error concealment of VVC video, showing how spatio-temporal neural priors can recover damaged decoded frames \cite{liu2024swinvec}. GRACE jointly trains neural video codecs under packet losses so that the codec itself becomes loss-resilient, reducing the need for retransmission in real-time communication \cite{cheng2024grace}. Reparo targets video conferencing and generates missing frame information conditioned on received content when Internet losses make retransmission or excessive FEC inefficient \cite{cheng2023reparo}. NERVE demonstrates that frame recovery, super-resolution, and enhancement-aware ABR can run on mobile devices in real time, which is essential for practical receiver-side deployment \cite{yeo2024nerve}. FrameCorr further treats incomplete transmitted data as a reconstruction problem under timing and bandwidth constraints \cite{li2024framecorr}. Together, these systems show that channel-aware reconstruction can operate at different levels, from codec-integrated robustness to device-side post-processing.

	Another line of work borrows mature video inpainting and temporal-consistency models as receiver-side recovery modules. Flow-guided video inpainting completes optical flow and propagates pixels across frames for temporal coherence \cite{xu2019deepflow}; E2FGVI unifies flow completion, feature propagation, and content generation in an end-to-end framework \cite{li2022e2fgvi}; and flow-guided Transformers use motion information to guide attention-based recovery \cite{zhang2022flowtransformer}. Other methods explicitly model temporal structure and alignment, such as spatial-temporal completion networks and temporal adaptive alignment networks \cite{wang2019stcn,wang2020stt}. \cite{lei2020dvp} further shows that temporal consistency can be enforced on a target video without relying on a large external training set. In communication scenarios, the loss mask, packet dependency, and channel state provide additional side information that can guide these reconstruction modules.
	
	\section{Large Model-Enabled Media Communication and Generative Transmission}

\begin{figure*}[htbp]
	\centering
	\includegraphics[width=6.7in]{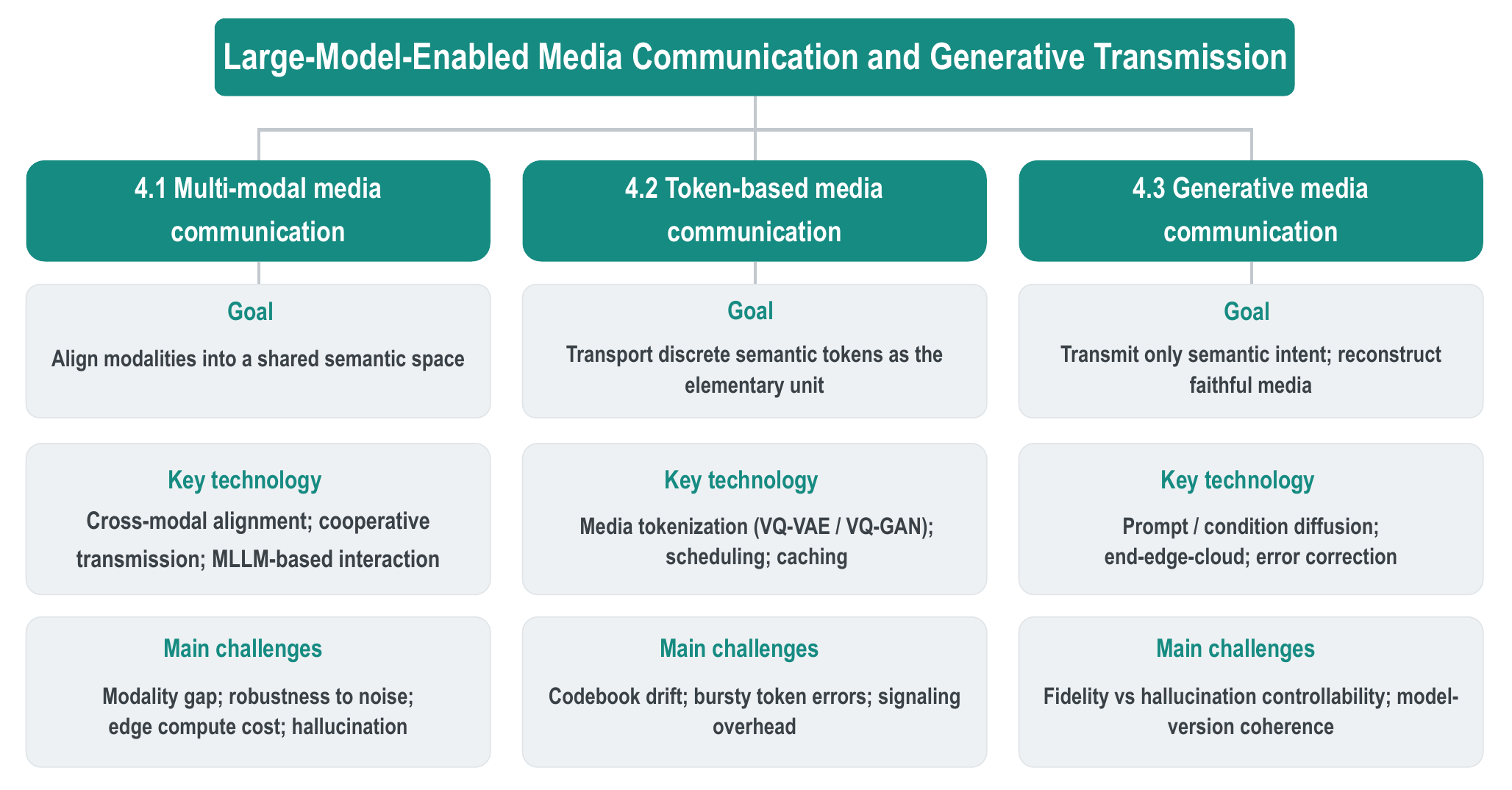}
	\caption{Overview of large model-enabled media communication and generative transmission.}
	\label{fig_8}
	\end{figure*}

	Previous sections have discussed AI-driven media processing and media-aware transmission. This section focuses on large model-enabled media transmission, where MLLMs, VLMs, and diffusion-based generative models provide multi-modal alignment, contextual understanding, instruction following, and content generation. These models allow heterogeneous modalities to be aligned in shared spaces~\cite{radford2021learning,sec4-2-li2023blip,sec4-3-girdhar2023imagebind,sec4-44-chen2024personalizing}, compact cues to drive reconstruction~\cite{rombach2022latentDiffusion,sec4-7-zhang2023adding}, and natural-language interaction to control media services~\cite{alayrac2022flamingo,liu2024visual}. From the communication perspective, they make tokens, prompts, latent features, and conditioning signals usable transmission units for local or edge-deployed reconstruction and enhancement~\cite{sec4-9-bourtsoulatze2019deep,sec4-10-qiao2025token,sec4-11-he2023rate,sec4-12-wu2024cddm,sec4-14-qiao2024latency,sec4-15-guo2025diffusion}. Accordingly, this section reviews multi-modal media communication, token-based media communication, and generative media communication, as shown in Fig. \ref{fig_8}.

	\subsection{Multi-modal Media Communication}
	Media in 6G is inherently multi-modal. Multi-modal media communication studies how heterogeneous media sources are represented, transmitted, and interpreted jointly rather than as isolated streams. Its technical basis is cross-modal structure: aligned representations enable modality substitution and transcoding, cooperative transmission allocates resources across modalities, and interactive models turn transmitted media into a queryable service interface. This subsection reviews these three capabilities.
	
	\subsubsection{Cross-modal Alignment}

	Cross-modal alignment aims to build a shared representation space where semantically related signals from different modalities are mapped close to each other, while unrelated signals are separated. This capability provides the foundation for large model-enabled media communication, because multi-modal substitution, token-level interoperability, and prompt-driven generation all require a common representation space. A representative milestone is CLIP~\cite{radford2021learning}, which demonstrates that natural-language supervision can learn transferable visual representations and support zero-shot recognition. Subsequent studies further extend cross-modal alignment in both scale and modality coverage. ALIGN~\cite{sec4-16-jia2021scaling} scales vision-language pretraining with large-scale noisy web data, while BLIP and BLIP-2~\cite{sec4-2-li2023blip} improve vision-language understanding by bridging frozen visual encoders and large language models. ImageBind~\cite{sec4-3-girdhar2023imagebind} further aligns images, text, audio, depth, thermal signals, and inertial measurements within a unified embedding space, showing the potential of multi-modal retrieval, composition, and generation.

	For 6G media communication, cross-modal alignment is particularly useful in the following aspects.

	\begin{itemize}
		\item \textbf{Modality interchangeability:} 
		Since different modalities can be projected into a shared embedding space, the network can flexibly select or substitute media forms according to channel conditions, device capability, and service requirements. For example, a textual description, a low-resolution thumbnail, or a depth map may provide complementary cues for reconstructing the same scene. This allows the system to transmit a more compact or robust modality when bandwidth is limited, while still preserving the essential content needed at the receiver.
	
		\item \textbf{Cross-modal transcoding:} 
		Aligned representations also enable intermediate nodes, such as base stations or edge servers, to translate information across modalities. For instance, audio can be converted into text, text can guide image generation, and visual content can be summarized into compact language descriptions. With deployed MLLMs and VLMs as backbones, such cross-modal transcoding can support adaptive media representation, edge-assisted reconstruction, and service-aware content delivery~\cite{alayrac2022flamingo,liu2024visual,sec4-13-xie2024toward}. This changes the role of the network from merely forwarding bitstreams to actively transforming media representations.
	
		\item \textbf{Language-assisted media control:} 
		VLMs, such as LLaVA~\cite{liu2024visual}, enable natural language to serve as a lightweight control signal for media communication. Compared with dense visual features or full-resolution frames, textual descriptions are compact and channel-friendly. They can be used to guide encoding decisions, transmission scheduling, ROI selection, and receiver-side reconstruction. In this way, language becomes not only a description of media content, but also a controllable interface between media understanding and wireless transmission.
	\end{itemize}

	Cross-modal alignment therefore links multi-modal media understanding with adaptive wireless media transmission. By mapping heterogeneous media into a shared semantic space, it enables modality substitution, cross-modal transcoding, and language-assisted control, allowing media information to be compressed, selected, translated, and reconstructed according to service requirements and network conditions.
	
	\subsubsection{Multi-modal Cooperative Transmission}

	Cross-modal alignment provides a shared semantic space for heterogeneous media, while multi-modal cooperative transmission further determines how different modalities should be jointly delivered over wireless links. In multi-modal services, images, videos, audio, text, depth, and sensor signals may describe the same scene or event from different perspectives. These modalities differ significantly in bitrate demand, distortion sensitivity, delay tolerance, and reconstruction value. Therefore, multi-modal cooperative transmission treats modality selection, bitrate allocation, channel coding, and physical-layer resource assignment as coupled optimization variables. The objective is to maximize end-to-end perceptual quality and task utility under dynamic channel conditions, computational constraints, and user requirements.

	According to available bandwidth and service requirements, multi-modal cooperative transmission can be divided into three representative regimes. In the bandwidth-rich regime, multiple modalities can be transmitted with relatively high fidelity, and the system mainly focuses on modality-aware compression, synchronization, and cross-modal fusion. In the bandwidth-limited regime, the system needs to selectively compress, simplify, or replace certain modalities. For example, a high-rate video stream may be substituted by a combination of low-resolution visual references, textual descriptions, semantic tokens, or generative prompts. In the bandwidth-collapse regime, only extremely compact information, such as text, keypoints, or token-level descriptions, can be transmitted, while the receiver relies on generative models to reconstruct missing modalities or visual details. The boundary among these regimes is not fixed, but depends on channel quality, latency constraints, terminal capability, and user intent.

	Representative studies have explored different forms of modality-aware transmission. A wireless semantic video conferencing system transmits sparse facial keypoints instead of full pixel frames and reconstructs facial expressions at the receiver, thereby significantly reducing transmission overhead for human-centric communication~\cite{sec4-21-jiang2022wireless}. A deep wireless video semantic transmission framework jointly optimizes learned video compression and channel coding, showing that source representation and wireless transmission can be trained together for robust video delivery~\cite{sec4-22-wang2022wireless}. For multi-modal services, a rate-adaptive coding scheme adjusts the coding rate of each modality according to channel signal-to-noise ratio (SNR) and content importance, enabling more flexible resource allocation across heterogeneous media streams~\cite{sec4-11-he2023rate}. These studies show that cooperative transmission is evolving from single-stream video optimization toward modality-aware and content-aware wireless delivery.

	Foundation-model-driven approaches further extend this paradigm by introducing semantic decomposition and generative reconstruction. A latency-aware generative semantic communication framework decomposes multi-modal content at the transmitter and assigns coding and modulation strategies according to user intent and modality importance~\cite{sec4-14-qiao2024latency}. A diffusion-driven generative communication framework combines a VAE-based compressor with a diffusion receiver, allowing high-quality reconstruction under severe bandwidth constraints~\cite{sec4-15-guo2025diffusion}. Compared with conventional multi-modal transmission, these methods do not require all modalities to be explicitly delivered. Instead, compact semantic cues, latent representations, or prompts can be transmitted, while missing visual details are reconstructed by receiver-side generative models.

	Multi-modal cooperative transmission extends media delivery from independent streams to joint modality-aware optimization. Coordinating modality selection, bitrate allocation, channel coding, and wireless resource assignment allows transmitted media forms to adapt to bandwidth, channel quality, terminal capability, and user intent, while foundation-model-driven decomposition and generative reconstruction reduce the need to deliver every modality explicitly.
	
	\subsubsection{Multi-modal Interaction Enhancement}

	Beyond bandwidth efficiency, large multi-modal models turn transmitted media into an interactive interface for querying, summarization, translation, annotation, and scene-level reasoning. From the communication perspective, the value of a delivered stream is determined not only by pixel-level fidelity, but also by the services that can be derived from it at the receiver, edge node, or cloud platform.

	Representative forms of multi-modal interaction enhancement can be summarized as follows:

\begin{itemize}
	\item \textbf{Real-time media enrichment:} 
	Large multi-modal models deployed at terminals or edge nodes can generate live captions, multilingual translations, event summaries, and auxiliary descriptions for transmitted audio-visual streams. These functions provide direct communication benefits. First, they improve robustness when the primary media stream is degraded, since captions or summaries can still preserve essential information under bandwidth shortage or packet loss. Second, they enhance accessibility for users with visual or hearing impairments without redesigning the source codec. Third, they allow the network to transmit complementary low-rate information together with the media stream, thereby improving the overall service experience under dynamic channel conditions.
	
	\item \textbf{Interactive media understanding:} 
	Multi-modal models enable users to issue content-related queries at the receiver, such as identifying important objects, summarizing recent events, locating defects, or extracting speaker-related information. Representative model families built upon vision-language understanding, such as BLIP-2~\cite{sec4-2-li2023blip} and LLaVA~\cite{liu2024visual}, provide the foundation for visual question answering, multi-modal dialogue, and agent-style scene parsing. For communication systems, this capability changes the transmission objective from delivering all visual details uniformly to delivering the information needed for downstream interaction. When the receiver can infer, retrieve, or summarize part of the content locally, the network may reduce redundant transmission while maintaining or even improving perceived utility.
	
	\item \textbf{Context-aware overlay and augmentation:} 
	In interactive and immersive services, large multi-modal and generative models can produce context-sensitive annotations, virtual objects, navigation cues, or interactive scene elements on top of the transmitted media stream~\cite{sec4-24-sehad2024generative,sec4-25-yang2024streamlined,sec4-26-jiang2024large}. These overlays can be generated according to user intent, scene context, and network conditions, rather than being fully transmitted as additional media layers. As a result, the communication system can transmit compact control information or semantic cues, while edge or terminal models generate the corresponding augmentation locally. This mechanism reduces transmission overhead and enables personalized interactive experiences.
\end{itemize}

	To support these interaction-enhanced services at the 6G scale, several system-level issues must be addressed.

\begin{itemize}
	\item \textbf{Low-latency inference and response:} 
	Interactive media services require tight response loops, especially in XR and remote collaboration scenarios. If captioning, querying, or overlay generation relies only on cloud-side inference, the additional round-trip latency may break real-time interaction. Therefore, LLM inference should be partially shifted to terminals or edge nodes through model compression, split inference, and collaborative end-edge-cloud deployment~\cite{sec4-19-lin2025pushing,sec4-20-lin2024split}.
	
	\item \textbf{Content-grounded interaction reliability:} 
	The outputs of multi-modal models, including answers, summaries, translations, and overlays, should remain grounded in the actually transmitted media content. Hallucinated descriptions or incorrect annotations may directly affect user trust and task reliability. This requirement couples interaction quality with source fidelity, channel reliability, and receiver-side model robustness. Therefore, communication systems need mechanisms for confidence estimation, source-content verification, and uncertainty-aware interaction.
	
	\item \textbf{Privacy-preserving query processing:} 
	User queries, interaction histories, gaze patterns, and contextual preferences may reveal sensitive personal information. When such information is offloaded to edge nodes or cloud platforms, privacy leakage becomes a major concern. Future systems should therefore combine local processing, encrypted transmission, access control, and privacy-preserving inference to protect both transmitted media and user interaction data.
	
	\item \textbf{Elastic communication-computing resource allocation:} 
	Unlike conventional media streaming, interactive workloads are bursty and user-driven. The required resources depend not only on media bitrate, but also on query frequency, model size, inference complexity, and interaction type. Therefore, 6G networks should jointly schedule bandwidth, computing, caching, and model-serving resources. Network slicing and edge resource orchestration should adapt to interaction demand rather than only to stream resolution or bitrate.
\end{itemize}

	Multi-modal interaction enhancement extends media communication from one-way content delivery to interactive services. By allowing transmitted media to be queried, interpreted, translated, summarized, and augmented in real time, large multi-modal models increase the service value of each delivered stream while making edge intelligence, user interaction, and privacy protection part of the communication design.

	\subsection{Token-based Media Communication}
\begin{figure*}[htbp]
	\centering
	\includegraphics[width=6.6in]{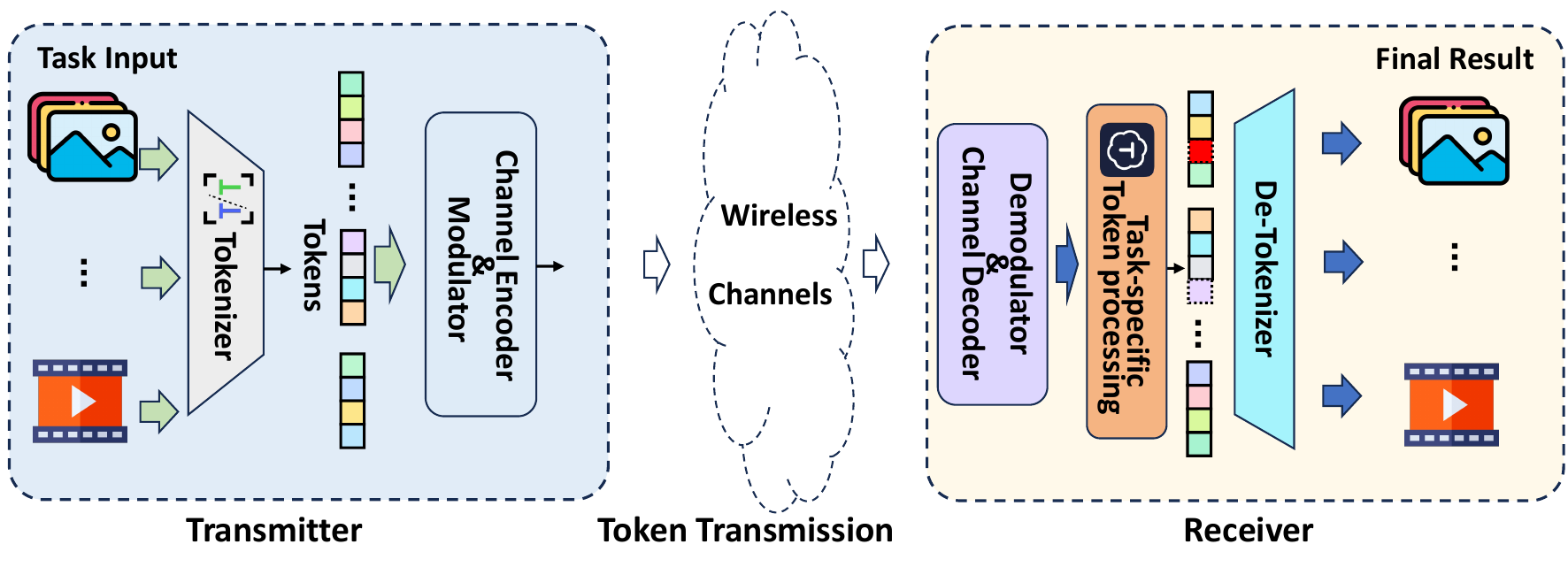}
	\caption{Token communication for wireless media transmission.}
	\label{fig_9}
	\end{figure*}

	The success of LLMs and VLMs has established tokens as the lingua franca of modern generative AI. Tokens are how text, images, video, and audio are now most often represented at the input and output of transformer-based models \cite{van2017vqvae,esser2021taming,sec4-29-chang2022maskgit,sec4-30-yu2024language,sec4-33-wang2026robustifying}, and the same property motivates a new communication paradigm in which semantic tokens become the logical units for scheduling, protection, caching, and reconstruction \cite{sec4-10-qiao2025token,sec4-13-xie2024toward,sec4-31-qiao2025todma,sec4-32-wei2025token}. The rationale is threefold. First, tokens are inherently compact and quantized, providing a natural digital interface to existing physical-layer modulation and coding stacks. Second, tokens carry interpretable semantic importance, where some tokens correspond to salient objects and others to background, which makes them natural scheduling units for differentiated transmission. Third, tokens are directly consumable by receiver-side foundation models, eliminating the lossy back-and-forth between continuous embeddings and discrete bitstreams that characterizes conventional pipelines. The illustration of token-based media communication is given in Fig. \ref{fig_9}. The remainder of this subsection covers how media is tokenized, how tokens are scheduled across the radio link, and how token-level caching reshapes content delivery at the edge.

	\subsubsection{Tokenization for Media Content}

	Media tokenization maps continuous image, video, or audio signals into sequences of discrete indices selected from a learned vocabulary. For communication systems, this operation provides a compact interface that converts high-dimensional media signals into transmittable, schedulable, and cacheable symbolic units. A representative foundation is the VQ-VAE ~\cite{van2017vqvae}, which encodes continuous features into discrete codebook entries and reconstructs the input from the corresponding indices. VQ-GAN~\cite{esser2021taming} further improves perceptual quality with adversarial and perceptual objectives, producing visual tokens with richer semantic and structural information. MaskGIT~\cite{sec4-29-chang2022maskgit} improves generation efficiency through parallel masked-token prediction, while video tokenizers such as MAGVIT and MAGVIT-v2~\cite{sec4-30-yu2024language} extend tokenization from images to spatio-temporal media. These studies show that discrete tokens can serve as compact media units for compression, generation, and transmission.

	From the communication perspective, tokenizer design directly affects transmission efficiency and reconstruction reliability. Codebook size and token granularity determine the trade-off among bitrate, fidelity, entropy-coding complexity, and error sensitivity. Finer tokens can preserve more details but require longer indices and stronger protection, whereas coarser tokens reduce bitrate and improve robustness at the cost of visual precision. Token semantic granularity is also important. Patch-level tokens are suitable for general compression, while object-level or scene-level tokens provide more interpretable units for importance-aware scheduling, unequal error protection, and semantic retransmission. In addition, modality-specific vocabularies are easier to optimize for individual media types, whereas unified multi-modal vocabularies can provide a common token interface for images, videos, audio, and text, consistent with token-based communication frameworks such as TokCom~\cite{sec4-10-qiao2025token}.

	Token-based transmission differs from continuous feature transmission. DeepJSCC transmit continuous learned features and can provide graceful degradation under channel noise~\cite{sec4-9-bourtsoulatze2019deep}, whereas token-based systems transmit discrete indices, making them naturally compatible with digital modulation, packetization, channel coding, caching, and retransmission. Discrete tokens are also more suitable for MLLM-based reasoning, symbolic manipulation, and cache-assisted media delivery. Therefore, media tokenization should not only be treated as a source coding tool, but also be jointly designed with channel coding, scheduling, and receiver-side reconstruction, so that token vocabularies, transmission protection, and generative decoding can be optimized as an integrated media communication system.
	
	\subsubsection{Token-level Transmission Priority Scheduling}

	Conventional wireless scheduling usually treats the transmitted bitstream as uniformly important and allocates physical-layer resources mainly according to channel quality, buffer state, and fairness constraints. However, when media content is represented as discrete tokens, different tokens may have highly unequal contributions to semantic fidelity, perceptual quality, and task utility. Tokens corresponding to foreground objects, facial expressions, motion boundaries, or task-relevant regions are usually more important than those describing smooth backgrounds or redundant textures. Therefore, token-level transmission priority scheduling connects media semantics with wireless resource allocation, allowing coding rate, MCS, retransmission, and protection level to be adjusted according to token importance.

	A key step is token importance estimation. Importance can be derived from saliency maps, attention weights, task objectives, or LLM-based contextual likelihoods. For example, TokCom ranks tokens according to MLLM-derived contextual importance~\cite{sec4-10-qiao2025token}, while intent-aware semantic decomposition assigns different weights to tokens according to task requirements~\cite{sec4-14-qiao2024latency}. In low-complexity scenarios, classical saliency or ROI models can also provide approximate importance indicators. Once the importance profile is obtained, the communication system can perform differential transmission. Important tokens can be assigned stronger channel coding, more reliable MCS, higher scheduling priority, or prioritized HARQ, while less important tokens may tolerate higher loss probability or lower protection. Related ideas in semantic communication and rate-adaptive multi-modal transmission~\cite{sec4-9-bourtsoulatze2019deep,sec4-11-he2023rate} can therefore be naturally extended to tokenized media streams.

	Recent studies further integrate token-level scheduling with multiple access and recovery. Token-Domain Multiple Access (ToDMA) treats tokens as elementary access units, where each device maps its tokenized source to a shared modulation codebook, and the receiver exploits token sparsity and codebook structure to separate collided transmissions~\cite{sec4-31-qiao2025todma}. This design suggests that token statistics can be used not only for compression, but also for multi-user access and interference resolution. In addition, context-aware iterative token detection combines channel observations with a pretrained masked language model to recover corrupted or missing tokens~\cite{sec4-32-wei2025token}. In this case, semantic priors are incorporated into the decoding process, linking token recovery with wireless reliability.

	Token-level transmission priority scheduling extends wireless resource allocation from packet-level management to semantic-unit-level optimization. By estimating token importance and mapping it to coding, modulation, retransmission, and access strategies, the system can protect the most valuable media information under limited wireless resources. Together with token-domain multiple access and context-aware token recovery, this paradigm connects LLM-enabled media representation with communication-aware resource orchestration.

	\subsubsection{Token Caching and Incremental Transmission}
	A practical observation underlies token caching: media tokens are spatially and temporally redundant across users, sessions, and scenes. Tokens that encode common objects (cars, faces, signage), scene templates (office, sidewalk, sky), and motion primitives recur far more often than their share of the codebook would suggest. If a copy of these high-frequency tokens is held at edge nodes, then the wireless link only needs to carry the incremental delta (the informative tokens of the current scene), drastically lowering repeated transmission cost. This idea is closely related to conventional content-delivery network (CDN) caching, but operates on semantic units that are inherently shareable across content rather than on byte-identical files.
	
	Several lines of recent work make this concrete. TokCom and its follow-ups \cite{sec4-10-qiao2025token,sec4-13-xie2024toward} explicitly contemplate context-aware repeat requests and routing in which intermediate nodes hold MLLM-derived priors over the token distribution and selectively retransmit only the tokens that the receiver's prior cannot fill in. Generative-AI-driven edge caching frameworks for XR and immersive media \cite{sec4-24-sehad2024generative,sec4-25-yang2024streamlined} pre-populate edge nodes with scene-template tokens and avatar embeddings keyed to anticipated user activity. 
	
	Token caching also brings new system challenges. Cache miss handling should degrade quality gracefully rather than collapse into a black frame, suggesting hybrid schemes in which an absent token is generatively filled in at the receiver. Tokenizer and model-version compatibility across heterogeneous edge nodes is critical: a token index is only meaningful in the context of the codebook that produced it. Storage-bandwidth-compute trade-offs need careful joint optimization, since edge storage, backhaul bandwidth, and edge compute are all finite and competing. Finally, privacy-preserving cache sharing is essential when user-specific tokens (faces, voices, locations) are involved, which links to the security-and-privacy research direction discussed later in this survey.
	
	\subsection{Generative Media Communication}
\begin{figure*}[htbp]
	\centering
	\includegraphics[width=6.6in]{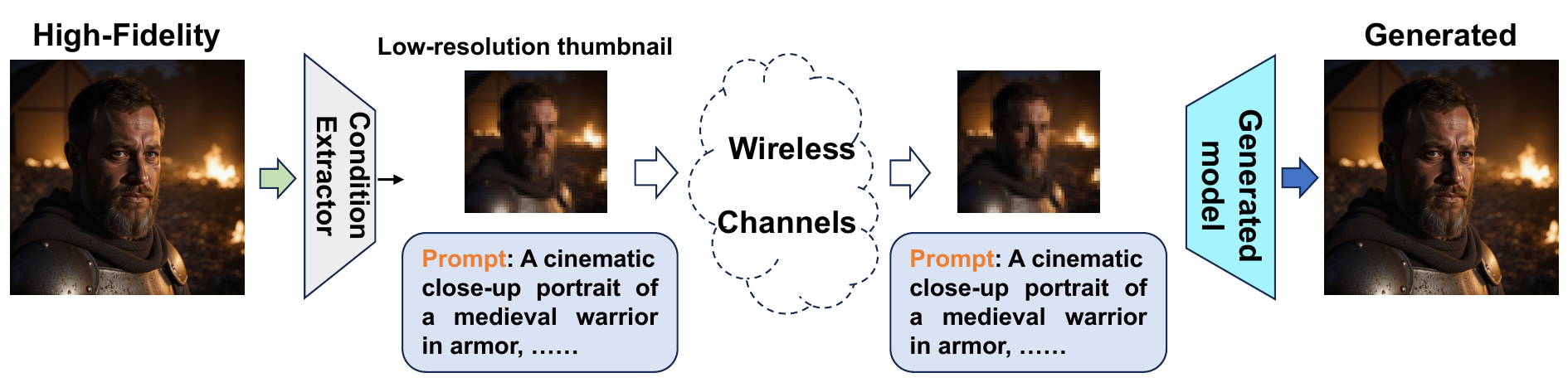}
	\caption{Generative media communication framework.}
	\label{fig_10}
	\end{figure*}

	Generative media communication uses receiver-side generative foundation models to synthesize or enhance media from compact conditions, partial observations, or tokenized representations \cite{sec4-9-bourtsoulatze2019deep,rombach2022latentDiffusion,sec4-7-zhang2023adding,sec4-14-qiao2024latency,sec4-15-guo2025diffusion}. The concept is shown in Fig. \ref{fig_10}. This subsection examines three angles: prompt-based generative transmission, end-edge-cloud collaborative generation, and generative error correction and quality enhancement.

	\subsubsection{Few-shot/Prompt-based Generative Transmission}

	Few-shot/prompt-based generative transmission provides a new transmission paradigm for bandwidth-constrained media services. Instead of sending dense pixel streams or high-dimensional features, the transmitter extracts compact generative conditions, such as natural-language descriptions, low-resolution references, structural maps, key tokens, or a small number of exemplar cues, and transmits them over the wireless link. The receiver then uses a pretrained generative model to reconstruct, enhance, or synthesize high-resolution media content from these conditions. In this paradigm, the communication payload is shifted from complete visual signals to compact prompts and conditioning information. This design can substantially reduce wireless transmission cost while maintaining acceptable perceptual quality for many media services, especially when exact pixel-level reconstruction is not required~\cite{rombach2022latentDiffusion,sec4-14-qiao2024latency}.

	The technical foundation of this paradigm is closely related to the development of diffusion-based generative models. Denoising Diffusion Probabilistic Models (DDPMs)~\cite{ho2020denoising} demonstrate strong distribution modeling capability for high-quality image generation. Latent Diffusion Models (LDM) and Stable Diffusion (SD) ~\cite{rombach2022latentDiffusion} further perform the denoising process in a compressed latent space, making high-resolution text-conditioned generation more efficient. ControlNet~\cite{sec4-7-zhang2023adding} introduces spatial conditioning, such as edge maps, human poses, segmentation maps, and depth maps, into pretrained diffusion backbones. These structured conditions are particularly suitable for communication scenarios, because they provide strong constraints on receiver-side generation while requiring much lower bitrate than full-resolution media streams.

	Recent studies have begun to integrate prompt-based generation with wireless transmission. A latency-aware generative semantic communication framework decomposes multi-modal content at the transmitter, encodes different semantic streams with appropriate coding and modulation strategies, and reconstructs visual content using a pretrained diffusion model at the receiver~\cite{sec4-14-qiao2024latency}. A diffusion-driven bandwidth-constrained communication framework combines a variational-autoencoder-based compressor with a diffusion receiver, enabling reconstruction under severe bandwidth limitations~\cite{sec4-15-guo2025diffusion}. Other diffusion-based approaches model channel impairment or channel denoising as part of the reconstruction process. For example, channel-denoising diffusion models introduce a learned denoising module after channel equalization~\cite{sec4-12-wu2024cddm}, while diffusion-enabled secure semantic communication further explores the robustness of generative reconstruction under eavesdropping or adversarial transmission conditions~\cite{sec4-42-he2026diffusion}. Diffusion-based generative transmission for joint denoising and bandwidth saving has also been extended to wireless video transmission~\cite{sec4-37-xie2025wireless}. These studies show that generative models can serve not only as media decoders, but also as receiver-side modules that jointly support denoising, reconstruction, and enhancement.

	The main trade-off in prompt-based generative transmission lies among prompt compactness, reconstruction fidelity, and source faithfulness. If the transmitted prompt is too short, the receiver may generate content that is semantically related but visually inconsistent with the source. If the prompt contains excessive detail, the bitrate advantage over conventional compression is reduced. Structured conditions, such as pose maps, depth maps, edge maps, segmentation masks, and low-resolution references, provide an effective compromise because they constrain the generative model more strongly than language alone while remaining relatively inexpensive to transmit. Therefore, the design of prompt representation should be jointly optimized with channel coding, modulation, retransmission, and receiver-side generation.
	
	Few-shot/prompt-based generative transmission reduces wireless payloads by sending compact generative conditions and using pretrained diffusion or latent-variable models at the receiver. Structured spatial conditioning, channel-aware denoising, and diffusion-based reconstruction have shown that robust visual communication is possible under severe bandwidth constraints, while also supporting personalized and interactive media services.

	\subsubsection{End-edge-cloud Collaborative Generation}

	Generative foundation models are often too computationally intensive for full on-device execution, while cloud-only inference may fail to satisfy the latency requirements of interactive media services. A practical 6G architecture should therefore distribute the generative pipeline across the end, edge, and cloud tiers by matching different model functions with their corresponding computing, latency, and freshness requirements~\cite{sec4-19-lin2025pushing,sec4-20-lin2024split,sec4-17-chen2024role}. This end-edge-cloud collaboration extends large-scale distributed inference to generative media transmission, where model execution, condition extraction, personalization, and content reconstruction need to be jointly coordinated.
	
	A representative partitioning can be organized as follows. The terminal performs lightweight operations, such as media capture, prompt extraction, user-intent encoding, and structured condition generation, including on-device pose or depth estimation. The edge executes the major part of generative inference, supports low-latency interaction, and hosts personalized adapters or lightweight model components for nearby users. The cloud is responsible for large model training, global knowledge updates, large-scale fine-tuning, and checkpoint management. Split inference and split learning provide practical mechanisms for this architecture~\cite{sec4-20-lin2024split}, where different parts of the model are deployed across tiers and intermediate activations or compact features are transmitted through radio and backhaul links. The partition point can be dynamically selected to balance inference latency, transmission overhead, and computing load. Adaptive layer splitting further formulates this decision as a model-based reinforcement-learning problem~\cite{sec4-19-lin2025pushing}, while edge-cloud offloading assisted by in-context reasoning shows that large models can also participate in their own deployment and resource orchestration~\cite{sec4-40-zhou2024generative}.
	
	This collaborative architecture indicates that generative media transmission should be co-designed with communication, computing, caching, and model management. Model compression techniques, such as distillation, quantization, and Mixture-of-Experts routing, can reduce inference cost at terminals and edge nodes, while energy-aware preprocessing determines how much semantic extraction should be performed on device. Meanwhile, tokenizer, adapter, and model-version consistency across distributed edge nodes is essential, since mismatched model components may cause semantic errors that cannot be detected by conventional packet-level checks. Therefore, end-edge-cloud collaborative generation provides a system-level foundation for scalable generative media services and naturally connects large model-enabled transmission with the AI-RAN and edge-media-service architectures discussed in Section~5.
	
	\subsubsection{Generative Error Correction and Quality Enhancement}

	The generative priors used for ultra-low bitrate transmission can also be employed for receiver-side error correction and quality enhancement, even when the transmitted media is produced by conventional codecs. In this setting, the generative model acts as an intelligent reconstruction module that compensates for residual transmission impairments, such as packet loss, dropped frames, low-resolution inputs, and motion blur. Instead of relying only on explicitly received symbols, the receiver combines available visual cues, channel observations, and learned media priors to recover missing or degraded content. Diffusion-based restoration has achieved rapid progress in super-resolution, inpainting, deblurring, and denoising, as summarized in recent surveys~\cite{sec4-41-li2025diffusion}, and these capabilities are largely built upon DDPM and latent diffusion backbones~\cite{ho2020denoising,rombach2022latentDiffusion}. This makes generative models particularly suitable for improving perceptual robustness in bandwidth-limited and error-prone wireless media transmission.

	Two representative usage patterns emerge in communication systems. The first is \textbf{generative error correction}, where the receiver uses an inpainting-style diffusion model, autoregressive model, or token recovery model to reconstruct regions corrupted by packet loss or quantization noise. Channel-denoising diffusion models jointly consider the wireless channel and diffusion denoising process for error correction~\cite{sec4-12-wu2024cddm}, while related diffusion-enabled designs extend this idea to secure semantic communication under potential eavesdropping~\cite{sec4-42-he2026diffusion}. For video transmission, decoupled diffusion-based multi-frame compensation further exploits temporal context to restore lost or degraded frames~\cite{sec4-37-xie2025wireless}. These methods indicate that generative priors can complement conventional channel decoding by recovering semantic and perceptual information that cannot be fully restored from corrupted bitstreams alone. The second usage pattern is \textbf{generative quality enhancement}, where the receiver applies generative restoration even when the bitstream is correctly decoded but has low visual quality due to bandwidth limitation or aggressive compression. Typical examples include super-resolution from low-rate video, deblurring under high mobility, and perceptual enhancement for low-bitrate streaming. Such enhancement modules do not necessarily replace classical FEC or retransmission mechanisms. Instead, they can be integrated into hybrid communication pipelines, where channel codes provide symbol-level reliability and generative models provide perceptual-level improvement. Recent work on conformal risk control further explores statistical guarantees for token communication systems, offering a promising direction for controlling reconstruction risk under uncertainty~\cite{sec4-33-wang2026robustifying}.

	Generative error correction and quality enhancement add receiver-side intelligence to 6G media communication. They augment channel decoding with perceptual reconstruction, combining symbol-level reliability with generative priors so that impaired or low-quality streams can be restored without abandoning conventional channel coding.
	
	\section{Intelligent Networks for Media Applications}
\begin{figure}[htbp]
		\centering
		\includegraphics[width=0.95\linewidth]{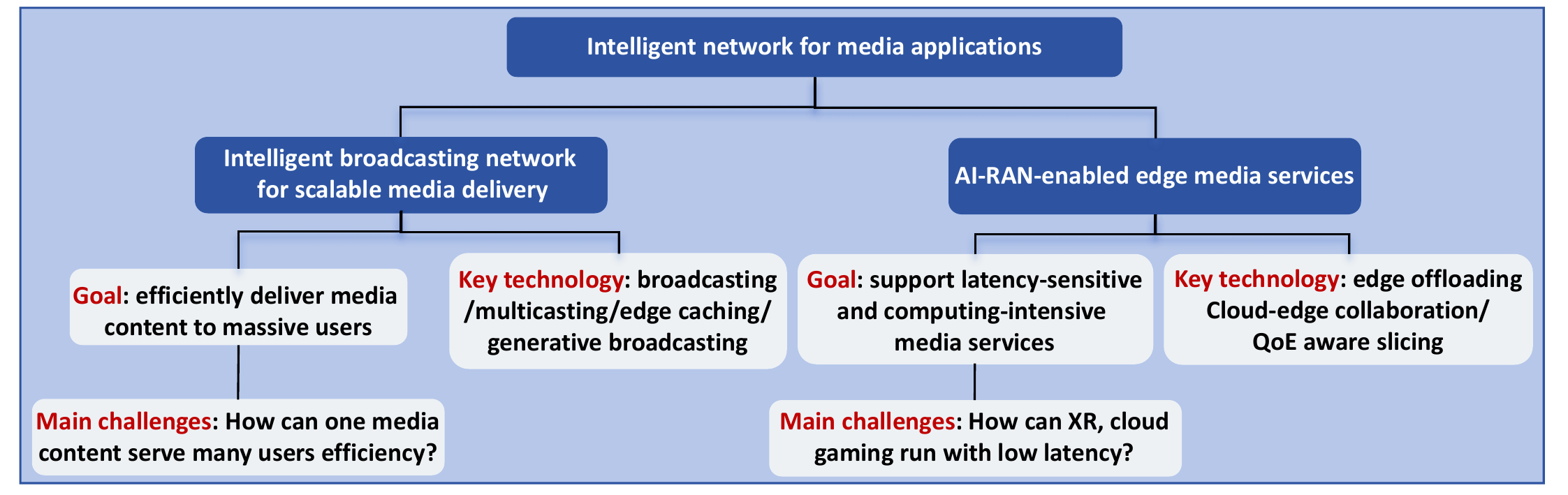}
		\caption{Overview of intelligent networks for media applications.}
		\label{fig11}
	\end{figure}

	This section discusses two representative directions: \textbf{intelligent broadcasting networks for scalable media delivery} and \textbf{AI-RAN-enabled edge media services}. The former focuses on exploiting broadcast/multicast capabilities, edge caching, proactive content push, air-ground-space coverage, and generative broadcasting mechanisms to improve media distribution efficiency. The latter emphasizes the integration of AI, radio access networks, and edge computing to support computation-intensive and latency-sensitive media applications. Together, these directions illustrate how future networks can evolve from connectivity-centric infrastructures to intelligent media service ecosystems. The main architecture is shown in Fig.~\ref{fig11}.
	
	\subsection{Intelligent Broadcasting Network for Scalable Media Delivery}
	Broadcasting has long been an efficient paradigm for delivering common content to a large number of users. Compared with unicast transmission, broadcast and multicast mechanisms can significantly reduce duplicated traffic when multiple users consume the same or similar media content. Nevertheless, conventional broadcasting systems are usually designed for one-way content dissemination and lack sufficient flexibility to adapt to heterogeneous user demands, dynamic network states, and personalized media experiences~\cite{broadcasting2}. With the emergence of 5G broadcast, multicast-broadcast services, edge computing, and AI-based content management, broadcasting networks are being redesigned as intelligent, adaptive, and service-aware infrastructures~\cite{5gbroadcast1}.
	
\begin{figure*}[htbp]
	\centering
	\includegraphics[width=6.7in]{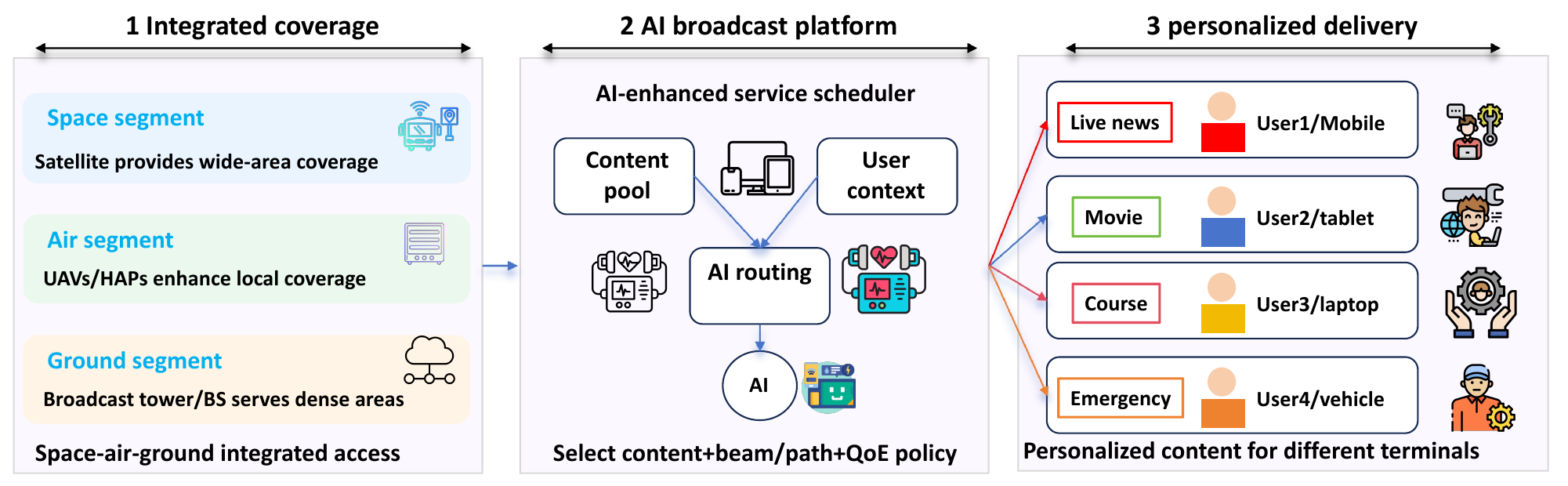}
	\caption{Illustration of intelligent broadcasting network.}
	\label{fig_12}
	\end{figure*}

	As shown in Fig. \ref{fig_12}, an intelligent broadcasting network integrates content awareness, user demand prediction, edge caching, multicast scheduling, and multi-domain coverage into a unified media delivery framework. Instead of transmitting all content reactively after user requests arrive, the network can predict popular content, pre-position media segments at edge nodes, and select suitable broadcast or multicast modes according to audience density and service requirements. For example, during a large-scale live sports event, a large portion of users may request the same high-definition stream, while others may require personalized camera angles, augmented statistics, or highlights. An intelligent broadcasting network can combine common-layer broadcast transmission with personalized unicast enhancement, thereby improving spectrum efficiency while maintaining service flexibility.
	
	The key challenge is to coordinate heterogeneous resources across the content, network, and computing domains. Media delivery performance depends not only on radio capacity but also on content popularity, cache placement, video encoding, user mobility, device capability, and edge processing availability. Therefore, intelligent broadcasting networks require cross-layer and cross-domain optimization. AI techniques can be used to predict content demand, infer audience distribution, identify congestion risks, and optimize content delivery strategies. Meanwhile, network slicing and programmable control can provide differentiated service guarantees for live broadcasting, video-on-demand, emergency media distribution, and immersive media services~\cite{network1}.

	\subsubsection{Edge Caching Coordinated with Broadcasting}
	Edge caching is a fundamental mechanism for improving media delivery efficiency. By storing popular media content close to users, edge caching reduces backhaul traffic, shortens delivery latency, and improves user experience. However, caching alone may not fully exploit the commonality of media requests. When many users request the same content within a short time window, broadcasting or multicasting the content from an edge node can further improve resource utilization~\cite{broadcasting5}. Therefore, coordinated edge caching and broadcasting is an important design principle for scalable media delivery.
	
	In a coordinated framework, caching decisions and broadcast scheduling are jointly optimized. The network first estimates content popularity and user distribution based on historical requests, social trends, location information, and event schedules. Popular media segments, such as live event previews, viral short videos, software-defined media assets, or frequently accessed video chunks, can be proactively cached at edge servers~\cite{video1}. When user demand exceeds a certain threshold, the network may switch from unicast delivery to multicast or broadcast delivery. In this way, the same cached content can be efficiently served to multiple users with reduced duplicated transmission.
	
	This coordination is particularly important for adaptive bitrate streaming. Modern video services usually divide content into segments encoded at multi-quality levels~\cite{video1}. Different users request different bitrates according to their device capability and channel condition. A naive caching strategy stores many redundant versions of the same content, while a naive broadcasting strategy transmits a quality level unsuitable for many users. Intelligent coordination can identify common base layers suitable for broadcast transmission and deliver enhancement layers through unicast or localized multicast~\cite{video2}. This layered delivery mechanism can balance scalability and personalization.
	
	Another important issue is cache replacement~\cite{cache1}. Media popularity is highly dynamic, especially for live events, breaking news, and social media trends. Static caching policies may quickly become inefficient. AI-based prediction models can be used to forecast short-term content popularity and update cache placement accordingly. Reinforcement learning can further optimize caching and broadcasting policies by interacting with network environments. The reward function may include cache hit ratio, transmission cost, latency, energy consumption, and QoE. In practice, deployment must consider model complexity, prediction uncertainty, and signaling overhead.
	
	\subsubsection{Dynamic Multicast and Proactive Push}
	Dynamic multicast is a flexible transmission mechanism that groups users according to their content demands, radio conditions, and service requirements. Unlike traditional broadcast, which delivers the same content to all users in a coverage area, dynamic multicast can create, modify, and release multicast groups according to real-time demand. This is particularly suitable for media services where user interests are partially overlapping but not identical~\cite{broadcasting7}. For instance, during a live concert, some users may watch the main camera view, while others may select close-up views, backstage views, or interactive augmented content. Dynamic multicast can serve each group efficiently while avoiding the inefficiency of pure unicast delivery.
	
	The effectiveness of dynamic multicast depends on accurate group formation and resource allocation. Users can be clustered based on requested content, delay tolerance, channel quality, mobility pattern, and device capability. The network then allocates radio resources and selects appropriate modulation, coding, and beamforming strategies for each multicast group. A conservative design may choose transmission parameters according to the weakest user in the group, which improves reliability but reduces spectral efficiency. Alternatively, hierarchical modulation and layered coding can be used to support heterogeneous users within the same multicast group. Proactive push further extends dynamic multicast by transmitting content before explicit user requests. The basic idea is that future media demands can often be predicted from context. For example, users near a stadium may soon request match highlights; commuters may request news clips during morning travel; viewers of an ongoing series may request the next episode after finishing the current one. By pushing likely requested content to edge caches or user devices in advance, the network can reduce peak traffic and improve perceived latency.
	
	However, proactive push introduces a tradeoff between efficiency and waste. If pushed content is not consumed, storage and transmission resources are wasted. Therefore, the network must evaluate the confidence of demand prediction and the cost of proactive delivery~\cite{push1}. AI models can estimate the probability of future content requests and select content for push based on expected utility. The utility may consider user preference, content size, deadline, cache capacity, radio resource availability, and energy consumption. In addition, user consent and privacy protection are essential, especially when proactive push relies on personal behavior prediction.
	
	Dynamic multicast and proactive push can be jointly optimized. For example, when the network predicts that a large group of users will request the same media content within a short time, it may proactively multicast the content to edge nodes or devices during off-peak periods~\cite{broadcasting9}. During peak demand, the content can then be served locally with minimal delay. This mechanism is highly suitable for time-sensitive media distribution, such as emergency alerts, live event highlights, educational broadcasting, and popular short video dissemination.
	
	\subsubsection{Air-Ground-Space Integrated Media Coverage Broadcasting}

	Future 6G media services are expected to provide continuous access in dense urban areas, maritime environments, and disaster zones. However, terrestrial networks alone may not always offer sufficient coverage, capacity, or resilience, especially when ground infrastructure is congested or difficult to deploy~\cite{terrestrial1}. Air-ground-space integrated networks, consisting of terrestrial base stations, unmanned aerial vehicles (UAVs), high-altitude platforms (HAPs), and satellites, provide a promising infrastructure for media coverage broadcasting. Different from common point-to-point media delivery, this architecture can distribute various media content (e.g. emergency information, live video, XR scene data, and cached media resources) to massive users over wide areas~\cite{broadcasting10}.

	In such an integrated broadcasting system, different network layers provide complementary functions for media services. Terrestrial networks support high-capacity and low-latency delivery in populated areas, making them suitable for local live streaming, interactive media, and personalized enhancement layers~\cite{terrestrial2}. UAVs can be rapidly deployed to collect aerial video, provide temporary coverage, or relay media streams in events and emergency scenarios. HAPs can support regional broadcasting with wider coverage than terrestrial base stations and lower latency than satellites, while satellites provide global reach for remote areas, maritime users, and large-scale content distribution. Through the coordination of these layers, media services can maintain continuity even when part of the ground infrastructure is unavailable or overloaded.

	This architecture is especially useful for broadcasting media content with common demands. For live sports, public concerts, emergency alerts, remote education, and public safety videos, satellites or HAPs can broadcast common base-layer content or popular media segments to a wide region, while terrestrial networks provide local retransmission, edge caching, or personalized enhancement. For XR and digital-twin services, common scene geometry, background textures, digital-twin states, semantic tokens, prompts, or generative model assets can be broadcast to edge nodes in advance, while user-specific viewpoints, interaction data, and high-quality enhancement layers are delivered through terrestrial or UAV-assisted links. This hierarchical broadcasting design reduces repeated unicast transmission and improves the scalability of media services.

	Several issues should be addressed to support air-ground-space integrated media broadcasting. First, media traffic should be routed, cached, transcoded, and broadcast according to latency, bandwidth, reliability, platform mobility, energy constraints, and service priority~\cite{terrestrial3}. Second, heterogeneous links across terrestrial, aerial, and satellite platforms require scalable and robust media transmission, including scalable video coding, layered media representation, FEC, adaptive bitrate control, and multicast/broadcast mechanisms~\cite{terrestrial4}. Third, AI-based controllers can predict link quality, user distribution, and content popularity, thereby optimizing coding rates, broadcast areas, cache placement, and routing strategies proactively~\cite{terrestrial5}. Therefore, air-ground-space integration is not only a coverage extension technique, but also a scalable and resilient broadcasting mechanism for future media services, including emergency communication, remote education, large-scale live broadcasting, XR, and digital twins~\cite{terrestrial6}.
	
	\subsubsection{Generative Broadcasting Network}
	The rise of generative artificial intelligence introduces a new paradigm for media production and delivery. In traditional broadcasting, media content is generated at the source and then transmitted to users. In a generative broadcasting network, part of the media content can be generated, personalized, or reconstructed at edge nodes or user devices based on compact semantic descriptions, prompts, models, or latent representations. This paradigm may significantly reduce transmission load and enable highly personalized media experiences~\cite{xu2024unleashing}.
	
	Generative broadcasting can be understood as a shift from bit-level content delivery to semantic-level content delivery~\cite{generative3}. Instead of broadcasting complete high-resolution media streams to all users, the network may broadcast common semantic information, scene descriptions, object representations, or generative model parameters. Edge servers or user devices can then synthesize customized media outputs according to user preferences, device capability, and context~\cite{generative4}. For example, a sports broadcasting service may transmit core event information and camera geometry, while different users generate personalized views, commentary styles, or highlight summaries. However, generative broadcasting raises several technical challenges. 
	
	\begin{itemize} 
		\item \textbf{Semantic representation and reconstruction quality:} The transmitted representation should be compact enough to reduce broadcasting overhead, while still sufficient to support faithful and consistent media generation. If the semantic description is too coarse, the receiver may generate visually plausible but source-inconsistent content. If the transmitted representation is too detailed, the bandwidth advantage over conventional broadcasting becomes limited. Therefore, semantic descriptors, prompts, tokens, latent features, and structural conditions should be carefully designed according to the required fidelity, latency, and personalization level. 
		\item \textbf{Communication-computing resource coordination:} Generative broadcasting reduces wireless transmission load by shifting part of the reconstruction burden to edge nodes or terminals. However, reducing transmitted bits usually increases computation demand for generative inference. Therefore, communication, computing, caching, and generation resources should be jointly optimized~\cite{xu2024unleashing}. When wireless bandwidth is scarce but edge computing is sufficient, compact prompts or semantic tokens can be broadcast and expanded locally. When edge resources are limited, the network may need to broadcast more explicit media layers to reduce generation complexity. 
		\item \textbf{Model distribution and version consistency:} Generative media services often rely on large models, codebooks, decoders, and asset libraries, which cannot be frequently transmitted to all users. The network therefore needs to cache model components at edge nodes, broadcast lightweight model updates, and manage version consistency among transmitters, edge servers, and receivers~\cite{semantic2}. This issue is particularly important for synchronized broadcasting, because the same prompt or token sequence may produce inconsistent outputs if different receivers use different model versions or decoding rules. 
		\item \textbf{Trustworthiness and content authenticity:} Since generative models can synthesize realistic media content, generative broadcasting must include mechanisms for provenance tracking, watermarking, content verification, access control, and misuse prevention~\cite{fernandez2023stable}. Inaccurate or manipulated generated content may have serious effects in broadcasting scenarios. Therefore, generated outputs should remain faithful to the semantic information, and the network should provide verification mechanisms to distinguish authentic, reconstructed, and modified content. 
	\end{itemize}
	
	Despite these challenges, generative broadcasting is a promising direction for future media networks. It can support personalized education, immersive entertainment, virtual events, interactive advertising, and real-time multilingual media services. More importantly, it changes the role of the network from content delivery to content intelligence~\cite{generative6}. The network no longer only transports media bits; it participates in deciding what information should be transmitted, where content should be generated, and how user experience should be optimized.

	\begin{figure*}[htbp]
	\centering
	\includegraphics[width=6.8in]{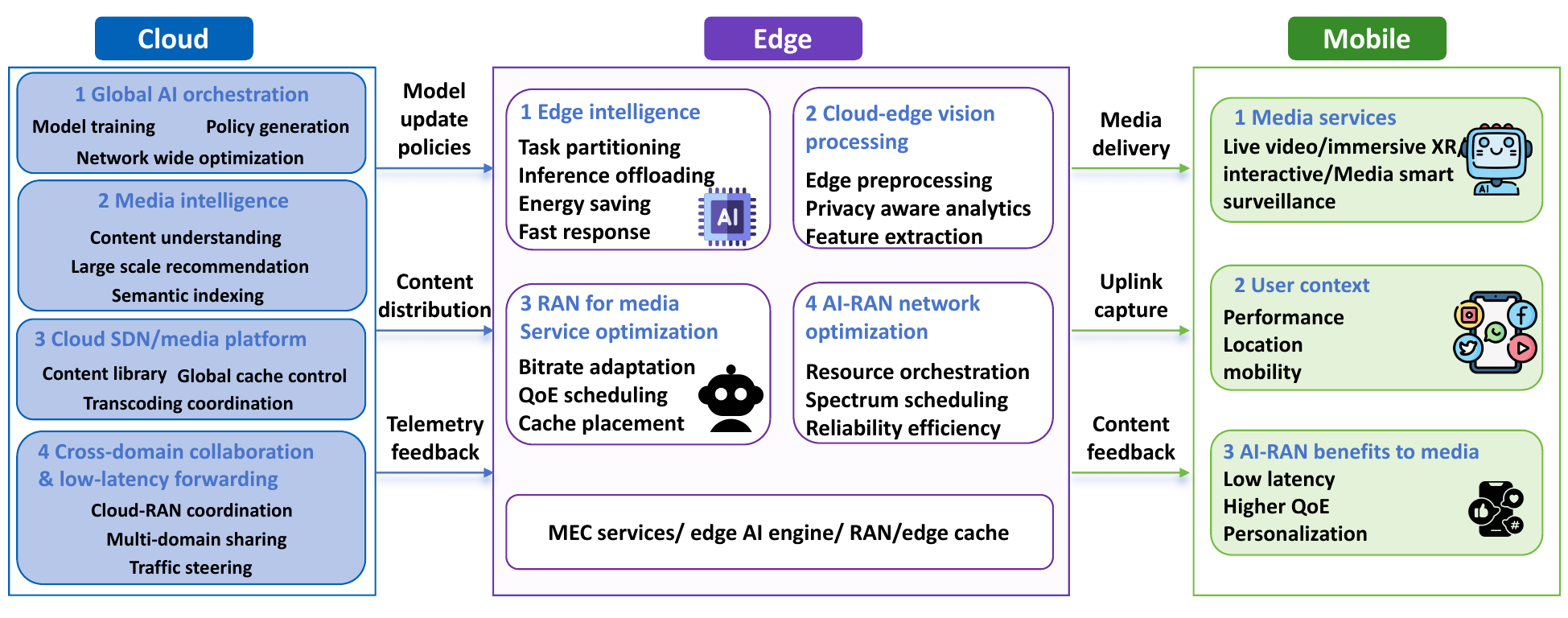}
	\caption{Illustration of AI-RAN for media services with cloud-edge-mobile cooperation.}
	\label{fig_13}
	\end{figure*}

	\subsection{AI-RAN-Enabled Edge Media Services}

	AI-RAN refers to the deep integration of AI and radio access networks, where AI is used not only as an application workload but also as a native control mechanism for radio resource management, mobility control, traffic prediction, network optimization, and service orchestration. For media applications, AI-RAN is important because many services are both communication-intensive and computation-intensive~\cite{RAN1}. Compared with traditional RAN architectures, AI-RAN provides a more service-aware and computation-aware control framework, enabling the network to infer media service requirements, estimate QoS and QoE, and jointly allocate radio, computing, caching, and AI inference resources~\cite{RAN2}. Edge nodes near base stations can support video enhancement, object recognition, scene understanding, content recommendation, and media transcoding, while radio transmission strategies can be adapted according to edge-computing status and media-flow requirements~\cite{RAN3}. Different from conventional mobile edge computing, AI-RAN emphasizes native integration with the radio access network and uses AI as both a media-processing engine and a network-optimization tool. The following subsections discuss representative functions, including edge intelligence offloading, cloud-edge vision processing, RAN optimization for media services, cross-domain collaboration, and network slicing for vision quality of service and quality of experience, as shown in Fig. \ref{fig_13}.
	
	\subsubsection{Edge Intelligence Offloading}
	Many media applications rely on AI inference or data analytics, such as object detection in video streams, super-resolution, background segmentation, gesture recognition, and scene understanding~\cite{EI1}. Executing these tasks entirely on user devices may be constrained by battery capacity, thermal limits, and hardware capability. Offloading them entirely to the cloud may introduce excessive latency and backhaul traffic. Edge intelligence offloading provides an intermediate solution by distributing AI workloads among devices, edge, and cloud \cite{EI2}.
	
	The core problem is to decide which computation tasks should be executed locally, at the edge, or in the cloud. For media services, this decision depends on several factors, including input data size, model complexity, latency requirement, wireless channel condition, edge server load, and energy consumption~\cite{EI3}. For example, a lightweight video enhancement model may run on the device, while a more complex multi-object tracking model may be offloaded to the edge. Long-term model training or global content analysis may still be performed in the cloud.
	
	AI-RAN can improve offloading decisions by jointly considering radio and computing states. Traditional offloading methods often focus on computing latency while assuming simplified communication models~\cite{EI4}. In contrast, media applications require fine-grained coordination between uplink transmission, edge inference, and downlink delivery. If the uplink channel is congested, offloading raw video frames may be inefficient. The device may instead perform feature extraction locally and offload compact intermediate representations. Conversely, if the edge server is overloaded, the network may allocate more radio resources for cloud offloading or shift some tasks back to devices~\cite{EI1}.
	
	Task partitioning is another important issue. Deep neural networks can be split across devices and edge servers. Early layers may process raw media data locally, while later layers perform high-level inference at the edge~\cite{EI6}. Such split inference can reduce transmission volume and preserve some degree of privacy. However, the optimal split point changes with channel quality, model architecture, and device capability. Dynamic partitioning algorithms are therefore needed. For real-time media applications, offloading reliability is as important as average latency. Occasional delay spikes may severely degrade user experience in interactive media services~\cite{media1}. Therefore, offloading policies should consider tail latency, jitter, and service continuity. Predictive AI models can forecast wireless channel variation and edge workload, enabling proactive migration or replication of AI tasks. Nevertheless, these mechanisms must be lightweight enough to operate within RAN control timescales.
	
	\subsubsection{Cloud-Edge Joint Vision Processing}
	Vision processing is a central workload in media applications that require real-time analysis of high-resolution video streams, creating heavy pressure on communication and computing resources~\cite{CE1}. Cloud-edge joint vision processing addresses this challenge by distributing vision tasks across multiple layers of the computing hierarchy.
	
	In a cloud-edge joint framework, edge nodes perform latency-sensitive processing, while cloud servers handle computation-intensive or global tasks~\cite{CE2,xie2023communication}. For instance, an edge server may conduct real-time object detection and tracking for a live video stream, while the cloud performs long-term behavior analysis, model retraining, and cross-camera correlation. This division allows the system to satisfy real-time requirements while benefiting from the large-scale computing capacity of the cloud.
	
	Media-aware feature compression is important in this framework~\cite{CE3}. Instead of transmitting raw video to the cloud, edge nodes can extract semantic features, object metadata, or event summaries. This can substantially reduce backhaul load while preserving information relevant to specific tasks~\cite{CE4}. For example, in sports broadcasting, edge nodes near cameras may extract player trajectories and ball positions, while the cloud generates tactical analysis and highlight clips. In smart city media services, edge nodes may detect abnormal events and only upload selected clips or metadata.
	
	Cloud-edge collaboration also supports adaptive model selection. Different vision models provide different tradeoffs among accuracy, latency, and computing cost. The system can select lightweight models during congestion and high-accuracy models when resources are sufficient. Moreover, multiple edge nodes can collaborate on multi-view vision tasks. For example, cameras deployed at different locations may capture the same scene from different angles. Joint processing can improve object tracking and scene reconstruction, but it requires synchronization, calibration, and efficient inter-edge communication~\cite{CE6}.
	
	A major challenge is maintaining consistency across distributed processing nodes~\cite{CE7}. If different edge nodes use different model versions or inconsistent metadata formats, the cloud may produce unreliable results. Therefore, model management, feature standardization, and synchronization protocols are necessary. Privacy is another concern, as video streams may contain sensitive personal information. Privacy-preserving vision processing, including on-device anonymization, edge-based filtering, federated learning, and secure execution environments, can reduce the risk of exposing raw media data.
	
	\subsubsection{RAN for Media Service Optimization}
	Radio access networks (RANs) play a direct role in determining media service quality. Bandwidth, latency, packet loss, jitter, handover performance, and scheduling fairness all affect user experience. Traditional RAN optimization focuses on network-centric metrics such as throughput, spectral efficiency, and coverage. However, media applications require service-centric optimization based on QoE. For example, a small reduction in throughput may be acceptable if video quality remains stable, while a brief interruption may be unacceptable for interactive gaming or live broadcasting.
	
	RAN for media service optimization aims to map media requirements into radio control decisions. This includes media-aware scheduling, adaptive bitrate coordination, mobility management, and congestion control~\cite{RAN11}. For video streaming, the RAN can cooperate with application-layer bitrate adaptation to prevent buffer starvation and reduce quality fluctuations. For XR, RAN must provide low-latency and high-reliability links to support motion-to-photon latency constraints. For live broadcasting, uplink reliability from cameras or mobile production units may be as important as downlink capacity to viewers.
	
	AI can help estimate user experience from network measurements and application indicators. Instead of optimizing throughput alone, the RAN can learn the relationship between radio conditions, media buffer status, video quality, and user-perceived experience~\cite{AIRAN1}. Based on this estimation, the scheduler can prioritize flows that are at risk of quality degradation. For example, a user with a nearly empty playback buffer may receive temporary resource priority, while a user with a sufficient buffer may tolerate lower instantaneous throughput.
	
	Another important direction is cross-layer feedback~\cite{cl1}. Media applications can expose service information to the network, such as frame deadlines, video layer importance, target bitrate, and latency sensitivity. The RAN can then treat different packets differently. For layered video, base-layer packets may receive higher priority than enhancement-layer packets because losing the base layer has more severe impact~\cite{cl2}. For real-time interaction, packets with expired deadlines may be dropped rather than transmitted late, thereby saving radio resources.
	
	Media service optimization must also consider fairness. Prioritizing premium or latency-sensitive services may degrade the experience of other users. Therefore, resource allocation should balance service differentiation, user fairness, and network efficiency. Network slicing provides a systematic mechanism for such differentiation, while AI-based control can dynamically adjust slice resources according to service demand.
	
	\subsubsection{Cross-domain Collaboration $\&$ Low-latency Forwarding}
	Edge media services involve multiple domains, including radio access, transport network, edge computing, cloud computing, content delivery networks, and application platforms. Optimizing only one domain may not guarantee end-to-end performance. For example, reducing radio latency is insufficient if edge processing is delayed, and increasing edge computing capacity is ineffective if the transport path is congested. Therefore, cross-domain collaboration is necessary for low-latency and reliable media delivery~\cite{cl3}.
	
	Cross-domain collaboration requires sharing abstracted state information among domains. The RAN may provide radio condition, mobility, and congestion indicators~\cite{cl4}. The edge platform may provide computing load, memory availability, and inference queue length. The transport network may provide path latency and packet loss information. The application layer may provide media deadlines, quality targets, and user interaction patterns. Based on this multi-domain information, the system can perform joint routing, task placement, and resource allocation.
	
	Low-latency forwarding is particularly important for interactive media. For interactive media services, user input must be transmitted to the processing node, rendered or processed, and returned to the user within a very short time~\cite{ll1}. Any unnecessary routing detour can degrade experience. Therefore, traffic should be steered to nearby edge nodes whenever possible. When users move, service continuity requires seamless migration of media sessions and AI tasks. This may involve state transfer, context replication, and predictive handover.
	
	Programmable networking technologies can support flexible low-latency forwarding. Software-defined networking can dynamically select paths according to service requirements~\cite{ll2}. Segment routing can steer traffic through specific edge nodes or network functions. In-network computing and programmable data planes may further reduce latency by processing selected media metadata within the network. However, these mechanisms must be carefully designed to avoid excessive control complexity.
	
	AI can assist cross-domain collaboration by predicting congestion, user mobility, and workload variation~\cite{AI1}. For example, before a large audience enters a stadium, the network may pre-allocate edge computing resources and adjust forwarding paths. During a live event, the system may detect rising demand for a specific camera angle and proactively cache or transcode the corresponding stream at nearby edge nodes. Such proactive collaboration can significantly improve scalability and responsiveness.
	
	\subsubsection{Network Slicing for Vision QoS/QoE}

	Network slicing provides isolated and customized logical networks over shared physical infrastructure, enabling differentiated support for services with heterogeneous requirements. For media applications, especially vision-oriented services, slicing is particularly important because different applications may require distinct combinations of bandwidth, latency, reliability, computing capability, storage, and privacy protection~\cite{nc1}. For example, public video streaming can tolerate several seconds of buffering, whereas remote control based on live video requires ultra-low latency, high reliability, and stable visual feedback. Therefore, network slicing offers a flexible mechanism to map diverse vision service requirements onto dedicated logical network resources.

	A vision QoS/QoE slice should include not only radio resources, but also edge computing, storage, AI models, and transport paths. Its configuration may specify target latency, minimum bitrate, frame-loss tolerance, inference accuracy, model execution deadline, privacy level, and security requirements. For smart broadcasting, one slice may support high-capacity downlink video distribution, while another slice supports low-latency uplink video contribution from cameras~\cite{nc2}. For augmented reality, the slice may jointly reserve uplink resources for sensor data, edge computing resources for rendering or inference, and downlink resources for visual feedback. Compared with conventional QoS provisioning, vision QoE-aware slicing is more complex because user experience depends not only on throughput and delay, but also on resolution, frame rate, rebuffering, visual distortion, quality switching, inference accuracy, and semantic consistency~\cite{nc3}.

	AI-based slice orchestration can dynamically adjust resources according to media service demand and network conditions. During normal periods, a media slice may operate with moderate radio and computing resources, while during a popular live event, the slice may expand to support massive concurrent users~\cite{nc5}. When edge computing load increases, the orchestrator can allocate additional accelerators, migrate inference tasks, or reduce model complexity. When radio resources become scarce, the system may prioritize base-layer video delivery, reduce enhancement-layer transmission, or adjust coding and rendering strategies. In this way, network slicing becomes a joint communication-computing-control mechanism rather than a static bandwidth reservation tool.

	Isolation, accountability, and reliability are also essential for vision QoS/QoE slicing. Multiple media services may share the same physical infrastructure while having different commercial, regulatory, or mission-critical requirements. Slicing can prevent resource contention among services, while monitoring mechanisms can verify whether service-level agreements are satisfied~\cite{nc6}. For emergency video communication, industrial monitoring, or remote operation, stronger isolation and reliability guarantees may be required to ensure stable media delivery. Network slicing for vision QoS/QoE thus bridges network capabilities and media service requirements, allowing future 6G networks to support heterogeneous vision applications in a flexible, scalable, and controllable manner.
	
	\section{Future Directions}
	
	Although vision communication has shown great potential in supporting efficient, intelligent, and personalized media services, practical 6G vision communication still faces several open challenges. 
	
	\begin{itemize}
		\item \textbf{Unified semantic representation and transmission protocols:} 
		Future vision communication requires unified semantic representations across modalities, scenarios, devices, and networks. Promising directions include visual token standards, object-level semantic descriptors, cross-modal alignment formats, and intent description languages. Built upon these representations, semantic-aware channel coding and transmission protocols should be developed to support efficient and interoperable visual information exchange. The objective is to achieve semantic-level compatibility and reduce the adaptation cost across heterogeneous vision communication systems.
		
		\item \textbf{End-edge-cloud collaborative generation and optimization:} 
		Generative models can significantly improve low-bitrate and personalized visual transmission, but their high computational cost makes direct deployment on terminals challenging. Future systems should exploit end-edge-cloud collaboration, where terminals perform lightweight processing, edge nodes support low-latency generation and enhancement, and cloud servers handle large-scale model training and updating. Key research issues include model partitioning, collaborative inference, dynamic task offloading, and resource-aware generation. The goal is to achieve high-fidelity visual output with low latency and acceptable terminal energy consumption.
		
		\item \textbf{Large model-driven adaptive vision services:} 
		Large models provide semantic understanding, reasoning, and multi-modal interaction capabilities for vision communication. They can analyze user preferences, device capabilities, and network conditions to adapt coding, transmission, and reconstruction strategies. This direction aims to enable personalized vision services in which resources are allocated on demand and visual quality is optimized according to user experience and task utility.
		
		\item \textbf{Security and privacy protection:} 
		Vision communication often involves sensitive visual information, such as faces, identities, locations, behaviors, and private environments. Future research should develop privacy-preserving visual representations and secure transmission mechanisms for visual tokens, semantic features, and generative prompts. In addition, generative vision communication introduces new risks, including hallucinated content, malicious manipulation, and fake media generation. Therefore, media and communication-level security techniques, such as content provenance, watermarking, authentication, physical-layer security, and tamper detection, should be integrated into the system design to ensure trustworthy and compliant visual communication.
		
		\item \textbf{Standardization and testbed validation:} 
		Practical deployment of 6G vision communication requires unified standards and reliable validation platforms. Future work should promote the standardization of semantic representations, system interfaces, transmission protocols, evaluation metrics, and network architectures. Meanwhile, multi-scenario testbeds should be constructed to evaluate representative vision services under realistic conditions. These testbeds are essential for evaluating end-to-end performance, latency, energy consumption, semantic consistency, and QoS/QoE, thereby accelerating industrial deployment and commercialization.
	\end{itemize}
	
	\section{Conclusions}
	
	This survey reviews the vision, key technologies, and opportunities of 6G vision communication. Unlike traditional systems that separate media processing and transmission, 6G aims to integrate vision and communication intelligence. With AI and large models, future systems will evolve toward unified frameworks that jointly optimize media representation, transmission, and reconstruction.
	
	Technically, 6G vision communication is supported by three pillars: AI-driven media processing, integrated media-communication transmission, and large model-enabled generative communication. These advances enable efficient coding, robust delivery, and semantic-aware, low-bitrate visual services, driving the shift from data transmission to intelligent media services.
	
	Future systems should be evaluated beyond bit-level accuracy, considering perceptual quality, semantic utility, adaptability, scalability, and trustworthiness. Progress relies on interdisciplinary advances across communication, AI, media, vision, networking, and security.
	
	Overall, 6G vision communication will support intelligent and immersive services by bridging physical and virtual worlds. With continued development, it is expected to move from concept to large-scale deployment.
	
	\bibliographystyle{scis}
	\bibliography{reference,media4Com}
	%\begin{thebibliography}{99}
	
	%\bibitem{1} Author A, Author B, Author C. Reference title. Journal, 2024, 38: 13--28
	
	%\bibitem{2} Author A, Author B, Author C, et al. Reference title. In: Proceedings of Conference, Place, 2024. 6--12
	
	%\end{thebibliography}
	
	%%%%%%%%%%%%%%%%%%%%%%%%%%%%%%%%%%%%%%%%%%%%%%%%%%%%%%%
	%%% Appendix sections. ¸½Â¼ÕÂ½Ú, ·Ç±ØÑ¡
	%%%%%%%%%%%%%%%%%%%%%%%%%%%%%%%%%%%%%%%%%%%%%%%%%%%%%%%
	%\begin{appendix}
	%\section{Name}
	
	%\end{appendix}
	
\end{document}